\documentclass{aa}
\usepackage{graphicx}
\usepackage[varg]{txfonts}
\usepackage{hyperref}
\usepackage{natbib,twoopt}
\bibpunct{(}{)}{;}{a}{}{,}             %% natbib format for A&A and ApJ

\begin{document} 
        
        \title{A multiwavelength study of the W33 Main ultracompact HII region} 
        
        \author{Sarwar Khan,
                \inst{1,3}
                Jagadheep D. Pandian
                \inst{1}
                \and
                Dharam V. Lal\inst{2}
                \and
                Michael R. Rugel\inst{3}
                \and
                Andreas Brunthaler\inst{3}
                \and
                Karl M. Menten\inst{3}
                \and
                F. Wyrowski\inst{3}
                \and
                S-N. X. Medina\inst{3}
                \and
                S. A. Dzib\inst{3,4}
                \and
                H. Nguyen\inst{3}
        }
        
        \institute{Indian Institute of Space Science and Technology (IIST), Trivandrum 695 547, India\\
                \email{skhan@mpifr-bonn.mpg.de}\\
                %         Indian Institute of Space Science and Technology (IIST), Trivandrum 695 547, India\\
                %         \email{jagadheep@iist.ac.in}
                \and
                National Centre for Radio Astrophysics - Tata Institute of Fundamental Research, Post Box 3, Ganeshkhind P.O., Pune 411007, India\\
                %       \email{dharam@ncra.tifr.res.in}
                \and
                Max-Planck-Institut f\"{u}r Radioastronomie, Auf dem H\"{u}gel 69, 53121 Bonn, Germany \\
                \and
                IRAM, 300 rue de la piscine, 38406 Saint Martin d'H\`eres, France\\
        }
        
        \date{Received ; accepted }
        
        % \abstract{}{}{}{}{} 
        % 5 {} token are mandatory
        
        \abstract
        {} %leave it empty if necessary  
        {The dynamics of ionized gas around the W33 Main ultracompact HII region is studied using observations of hydrogen radio recombination lines and a detailed multiwavelength characterization of the massive star-forming region W33 Main is performed.}
        {We used the Giant Meterwave Radio Telescope (GMRT) to observe the H167$\alpha$ recombination line at 1.4~GHz at an angular resolution of 10$\arcsec$, and \textit{Karl. G. Jansky} Very Large Array (VLA) data acquired in the GLOSTAR survey that stacks six recombination lines from 4$-$8~GHz at 25$\arcsec$ resolution to study the dynamics of ionized gas. We also observed the radio continuum at 1.4~GHz and 610~MHz with the GMRT and used GLOSTAR 4$-$8~GHz continuum data to characterize the nature of the radio emission. In addition, archival data from submillimeter to near-infrared wavelengths were used to study the dust emission and identify young stellar objects in the W33 Main star-forming region.}
        {The radio recombination lines were detected at good signal to noise in the GLOSTAR data, while the H167$\alpha$ radio recombination line was marginally detected with the GMRT. The spectral index of radio emission in the region determined from GMRT and GLOSTAR shows the emission to be thermal in the entire region. Along with W33 Main, an arc-shaped diffuse continuum source, G12.81$-$0.22, was detected with the GMRT data. The GLOSTAR recombination line data reveal a velocity gradient across W33 Main and G12.81$-$0.22. The electron temperature is found to be 6343~K and 4843~K in W33 Main and G12.81$-$0.22, respectively. The physical properties of the W33 Main molecular clump were derived by modeling the dust emission using data from the ATLASGAL and Hi-GAL surveys and they are consistent with the region being a relatively evolved site of massive star formation. The gas dynamics and physical properties of G12.81$-$0.22 are consistent with the HII region being in an evolved phase and its expansion on account of the pressure difference is slowing down.}
        {}
        
        \authorrunning{S. khan et al.}
        \titlerunning{Multiwavelength study of W33 Main region}
        \maketitle

        \section{Introduction} \label{sec:intro}
        High-mass stars play an important role in the dynamics and evolution of the interstellar medium and the Galaxy. Along with supplying a large amount of ultraviolet (UV) photons to their surroundings, they also provide a feedback mechanism for the formation of the next generation of stars in a sequential manner \citep{1977ApJ...214..725E, 1987IAUS..115....1L}. The UV photons ionize the surrounding interstellar medium leading to the creation of HII regions. The HII regions may retain significant quantities of dust which can absorb an average of 34\% of the UV photons~\citep{2018ApJ...864..136}, re-emitting them at infrared wavelengths. While a massive star initially creates a hypercompact HII region with a typical size $\lesssim$ 0.05~pc, which expands to an ultracompact HII region with a size $\lesssim$ 0.1~pc, the UV radiation from multiple massive stars may lead to the formation of a classical HII region complex in which the individual HII regions overlap with one another. These regions are bright at radio frequencies due to thermal bremsstrahlung emission. The study of these regions at radio wavelengths gives insight into the properties of the ionized gas. The physical conditions and the dynamics in HII regions can be probed by studying their emitted radio recombination lines (RRLs). In contrast, studies of HII regions at near-infrared wavelengths provide insight on the central stellar population, in particular on the massive young stellar objects (YSOs) they harbor. Thus, a multiwavelength study of HII regions can provide a comprehensive picture of the star formation activity in a massive star-forming region and their surroundings.
        
        W33 is a massive star-forming region consisting of many HII regions, OB stars, and massive dust clumps in different evolutionary stages \citep{2014A&A...572A..63I}. \citet{1984ApJ...283..573S} conducted multi-band far-infrared (20-250~$\mu$m) observations of the W33 complex and detected four distinct far-infrared sources in the complex, namely W33 Main, W33 A, W33 B, and W33 B1. Dust emission at 870~$\mu$m (ATLASGAL; ~\citealt{atlasgal}) from the W33 complex reveals three large complexes -- W33 A, W33 B, and W33 Main -- along with several smaller clouds such as W33 A1, W33 B1, W33 Main1, among others. In this paper, we study the W33 Main region which hosts the brightest radio source in the W33 star-forming complex. The trigonometric parallactic distance to W33 determined from water masers is 2.4$^{+0.17}_{-0.15}$~kpc, which suggests an assignation to the Scutum spiral arm~\citep{2013A&A...553A.117I}. The total bolometric luminosity and mass of the W33 complex are $ 5.5\times 10^5$~L$_\sun$ and $1.02 \times 10^4$~M$_\sun$, respectively \citep{2014A&A...572A..63I}. 
        
        W33 Main was found to have an enhanced ratio of the CO (3$-$2) to (1$-$0) line flux (R$_{3-2/1-0} > 1.0$; \citealt{2018PASJ...70S..50K}). This was interpreted as an indicator of the presence of massive stars since outflows and the strong radiation from massive stars heat the surrounding molecular clouds, resulting in a high value of R$_{3-2/1-0}$. \citet{messineo} performed a near-infrared spectroscopic survey of bright stars in selected regions of W33, which led to the identification of 14 early-type stars (OB and Wolf-Rayet type) with ages consistent with $\sim 2-4$~Myr.
        
        Observations of H$_2$CO and recombination lines toward the W33 complex reveal the presence of two velocity components at 35~km~s$^{-1}$ and 60~km~s$^{-1}$ which are associated with W33 Main and W33~B, respectively \citep{1978A&A....65..307G,1978A&A....64..341B}. Similarly, \citet{2018PASJ...70S..50K} studied the molecular cloud of W33 using different CO lines and detected three velocity components at 35~km~s$^{-1}$, 45~km~s$^{-1}$, and 58~km~s$^{-1}$. Based on morphological correspondence and the fact that their R$_{3-2/1-0}$ was greater than one, both the 35~km~s$^{-1}$ and the 58~km~s$^{-1}$ components were suggested to be associated with W33. In contrast, the association of the 45~km~s$^{-1}$ component with W33 is still not clear. \citet{2013A&A...553A.117I} found that water masers in W33~A, W33 Main (velocity of 35~km~s$^{-1}$), and W33~B (velocity of 58~km~s$^{-1}$) have the same parallactic distance which suggests that all of their molecular clumps belong to the W33 massive star-forming complex. \citet{2021A&A...646A.137L} studied the large-scale environment around W33 using the rotational (1$-$0) transitions of $^{12}$CO, $^{13}$CO, and C$^{18}$O from the Purple Mountain Observatory and find evidence for a hub-filament system, with W33 being located at the central hub. \citet{2020MNRAS.496.1278D} also suggest that the star formation activity in W33 is triggered by converging and colliding flows from two velocity components $\sim 35$ and 53~km~s$^{-1}$.
        
        \citet{2014A&A...572A..63I} constructed spectral energy distributions (SEDs) of the different clumps using data from the MSX, Hi-GAL, and ATLASGAL surveys. The SEDs were fit using a two component model from which the temperature of the cold dust component, peak H$_2$ column density, and clump mass were determined. The temperature of the cold dust was found to vary between 25.0 and 42.5~K, suggesting that the clumps are in different evolutionary stages.
        
        In this paper, we report the results of high resolution observations of the RRL   and radio continuum at 1400 and 610~MHz, respectively, using the Giant Metrewave Radio Telescope (GMRT; \citealt{swarup:20}) and continuum and RRL data from the GLOSTAR survey \citep{2021A&A...651A..85B} toward W33 Main. Also, based on data taken at submillimeter to near-infrared wavelengths, we present an analysis of the emission from the cool dust and the YSOs.
        
        \section{Observations and data analysis} \label{sec:observation}
        
        \subsection{GMRT radio observations}
        \begin{table}[h!]
                \caption{Detail of the GMRT observation.}
                \label{tab:obs_detail}
                \centering
                \begin{tabular}{c c c }         \hline \hline
                        Frequency (MHz) & 1400 & 610 \\
                        \hline
                        Date of Observation & 23 Aug 2016 & 16 Sep 2016  \\
                        On source time (h) & $\sim$ 3 & 0.5 \\
                        Bandwidth (MHz) & 16 & 32\\
                        Number of Channels & 512 & 256 \\
                        Velocity Resolution (km s$^{-1}$) & 6.7 & $\ldots$ \\
                        Target RRL & H167$\alpha$ & $\ldots$ \\
                        RRL rest Frequency (MHz) & 1399.368 & $\ldots$ \\ 
                        Primary beam ($\arcmin$) & 26.2 & 44.4 \\
                        Synthesized beam ($\arcsec$) & 4.1 $\times$ 2.14 & 8.25 $\times$ 4.17 \\
                        \hline
                \end{tabular}
        \end{table}
        The 1400~MHz observations were carried out with the GMRT on 2016 August 23. The observations covered the H167$\alpha$ radio recombination line that has a rest frequency of 1399.368~MHz. The phase center for all the GMRT observation was 18$^h$14$^m$13$\fs$96 and $-$17$\degr$55$\arcmin$44$\farcs$87. The GSB backend was used with a bandwidth of 16 MHz that was split into 512 channels giving a velocity resolution of 6.7 km~s$^{-1}$. The radio sources 3C48 and 3C286 were used as flux and bandpass calibrators, while 1911$-$201 was used as the gain calibrator. The total observing time was 4.6~h with an on-source integration time of approximately 3~h. The 610~MHz observations were carried out on 2016 September 16 with the GSB configured to have a bandwidth of 32 MHz and 256 channels. The sources 3C48 and 3C286 were used as primary calibrators, while 1911$-$201 and 1743$-$038 were used as gain calibrators. The W33 Main position was observed with an integration time of 30 minutes. The primary beam of a GMRT antenna at 1400~MHz and 610~MHz are $24\arcmin$ and $43\arcmin$, respectively. The details of the observations are listed in Table~\ref{tab:obs_detail}.
        
        The data were reduced using the Astronomical Imaging and Processing Software (AIPS). The flagging of bad data and radio frequency interference and subsequent calibration were done using standard procedures. The continuum map of the region was then produced by imaging the line-free channels. Since GMRT does not carry out Doppler tracking, the spectral line data were first aligned using the task \texttt{CVEL}. The spectral line was found to be very weak with the line strength comparable to the residual ripple in the passband following bandpass calibration. However, the wavelength of the ripple was much larger then expected line width of the recombination line. Hence, the continuum was subtracted in a small sub-band of 32 channels around the line using the task \texttt{UVLSF}. The continuum subtracted data were then cleaned using standard procedures. In order to improve the signal-to-noise ratio (S/N), the spectral line cube was imaged using a uv taper of 50~k$\lambda$ and natural weighting. The spectral line cube was seen to have stripe artifacts due to bad data from certain baselines. These data were identified and removed using the tasks \texttt{UVMOD} and \texttt{UVSUB}. The final angular resolution of the spectral cube is $10\arcsec$, while that of the 1400 and 610~MHz continuum maps are $4.1\arcsec \times 2.1\arcsec$ and $8.3\arcsec \times 4.2\arcsec$, respectively. The data were then self-calibrated in order to reduce the residual calibration errors and improve the S/N.
        
        The amplitudes of the 1400 and 610~MHz data have to be corrected for the contribution of the Galactic plane to the system temperature, which is not accounted for in the calibration process due to the calibrators being located outside the Galactic plane \citep{2004MNRAS.349L..25R}. The amplitudes were hence scaled by a correction factor of $(T_\mathrm{gal} + T_\mathrm{sys})/T_\mathrm{sys}$, where $T_\mathrm{sys}$ is the system temperatures corresponding to the calibrators and $T_\mathrm{gal}$ is the contribution of the Galaxy, which was estimated using the 408~MHz map of \citet{haslam} and a spectral index of $-2.6$, where the flux density is defined as $S_\nu$ $\propto$ $\nu^\alpha$. The correction factors at 1400~MHz and 610~MHz were found to be 1.26 and 2.65, respectively. The data were then corrected for the response of the primary beam. The final root mean square (rms) in the radio maps were 1.6, 0.91, and 5.6~mJy~beam$^{-1}$ for the spectral cube, 1400~MHz continuum, and 610~MHz continuum, respectively.
        
        \subsection{The GLObal view on STAR Formation in the Milky Way (GLOSTAR) VLA data}
        GLOSTAR is a survey of the Galactic Plane from 4$-$8~GHz covering 145 square degrees from 358$\degr \leq$ $\ell$ $\leq$ 60$\degr$, |b| $\leq$ 1$\degr$, and the Cygnus~X region. The survey was carried out using the \textit{Karl G. Jansky} Very Large Array (VLA) in the compact D configuration and the more extended B configuration. The data products from the survey include continuum emission from 4.2$-$5.2~GHz and 6.4$-$7.4~GHz in full polarization, the H$_2$CO line at 4.83~GHz, methanol maser line at 6.7~GHz, and seven radio recombination lines of hydrogen. The full technical details and data reduction of the survey are described in \citet{2021A&A...651A..85B}.
        
        We obtained the GLOSTAR VLA-D configuration continuum image and the recombination line data cube. The continuum emission image is at a reference frequency of 5.8~GHz and has been constructed from mosaic of images of several frequency bins between 4 and 8~GHz (~\citealt{2019A&A...627A.175M, 2021A&A...651A..85B}; Dzib et al., in prep.). The D-configuration continuum images are restored with a circular beam of 18$\arcsec$. We use six radio recombination lines observed in the GLOSTAR survey (H98$\alpha$, H99$\alpha$, H110$\alpha$, H112$\alpha$, H113$\alpha$, H114$\alpha$), which are stacked in velocity in order to increase the S/N. The data were gridded and imaged at a spectral resolution of 5~km~s$^{-1}$ and all the line cubes were smoothed to a spatial resolution of 25$\arcsec$ before stacking. The full description of the data reduction and calibration of the RRLs is given in~\cite{2021A&A...651A..85B} and Rugel et al. (in prep).
        
        \subsection{Analysis of far-infrared data}
        
        In order to examine the cool emission toward W33 Main, we used data from the ATLASGAL \citep{atlasgal} and Hi-GAL \citep{molinari:10} surveys. We modeled the dust emission by fitting the SED from 870 to 70~$\mu$m on a per pixel basis. Since the images at different wavelengths have different resolution, plate scales, and data units, we first processed the data to have uniform resolution, plate scales, and data units. This analysis was carried out using the Herschel Interactive Processing Environment (HIPE). The data unit was converted to Jy~pixel$^{-1}$ for all wavelengths using the task ``ConvertImageUnit''. Next, the task ``PhotometricConvolution'' was used to project the data to a common grid with the resolution and plate scale of the 500~$\mu$m map since it has the poorest resolution of all bands. The premade kernel~\citep{2011PASP..123.1218A} was used for projecting the Herschel data, while the Gaussian kernel was used for the ATLASGAL data. After photometric convolution, all maps have the same resolution of 37$\arcsec$ and a plate scale of 7$\arcsec$~pixel$^{-1}$. The maps show emission from the source as well as foreground and background emission from the Milky Way. After subtracting a constant background, the emission from each pixel was modeled by a gray body as given in eq. (\ref{grey}):
        \begin{equation}
        \label{grey}
        S_\nu - I_{\nu,bg} = \Omega B_\nu(T_d) (1-e^{-\tau_\nu})
        ,\end{equation} 
        where $S_\nu$ is the flux density of each pixel, $I_{\nu,bkg}$ is the estimated background, $\Omega$ is the solid angle of an individual pixel ($7\arcsec \times 7\arcsec$), $B_\nu$ is the Planck blackbody function, $T_d$ is the dust temperature, and $\tau_\nu$ is the optical depth, which depends on the frequency as $\tau_\nu = \tau_0(\nu/\nu_0)^\beta$, where $\nu_0$ is a reference frequency (chosen to correspond to a wavelength of 500~$\mu$m) and $\beta$ is the dust emissivity. The SED fits were carried out using the ``MPFIT'' function of Python's ``PYSPECKIT'' module keeping $T_d$, $\tau_0$, and $\beta$ as free parameters with $\beta$ being constrained to vary between one and three.
        
        \subsection{Ancillary data}
        
        Our study was complemented with data at mid-infrared and near-infrared wavelengths. We used data from the \textit{Spitzer} GLIMPSE survey ($3.6-8$~$\mu$m; \citealt{2003PASP..115..953B}), the UKIDSS Galactic Plane survey ($1.25-2.15$~$\mu$m; \citealt{ukidss:57}), and 2MASS ($1.25-2.15$~$\mu$m; \citealt{2006AJ....131.1163S}) to study the YSO population and the ionizing stars in the region.
        
        \section{Results and discussion}
        \subsection{Radio continuum emission}
        \begin{table}[h!]
                \caption{Summary of radio continuum emission toward W33 Main at different frequencies.}
                \label{tab:emission_summ}
                \centering
                \begin{tabular}{c c c c}
                        \hline \hline
                        Parameters &  \multicolumn{2}{c}{GMRT}  & GLOSTAR  \\
                        \hline
                        Frequency (MHz) & 1400 & 610 & 5799.10 \\
                        Noise (mJy beam$^{-1}$) & 0.91 & 5.6 & 1.2\\ \hline
                        Peak flux (Jy beam$^{-1}$) &&&\\ \hline
                        W33 Main& 0.17 & 0.12 & 10.0 \\
                        G12.81$-$0.20&0.02&0.08&0.31 \\ \hline
                        Integrated flux (Jy)&&&\\ \hline
                        W33 Main &9.82&3.2&24.8\\
                        G12.81$-$0.20&2.86&3.57& 2.3\\
                        \hline
                \end{tabular}
        \end{table}
        
\begin{figure*}[h!]
                \centering
                \includegraphics[scale=0.45]{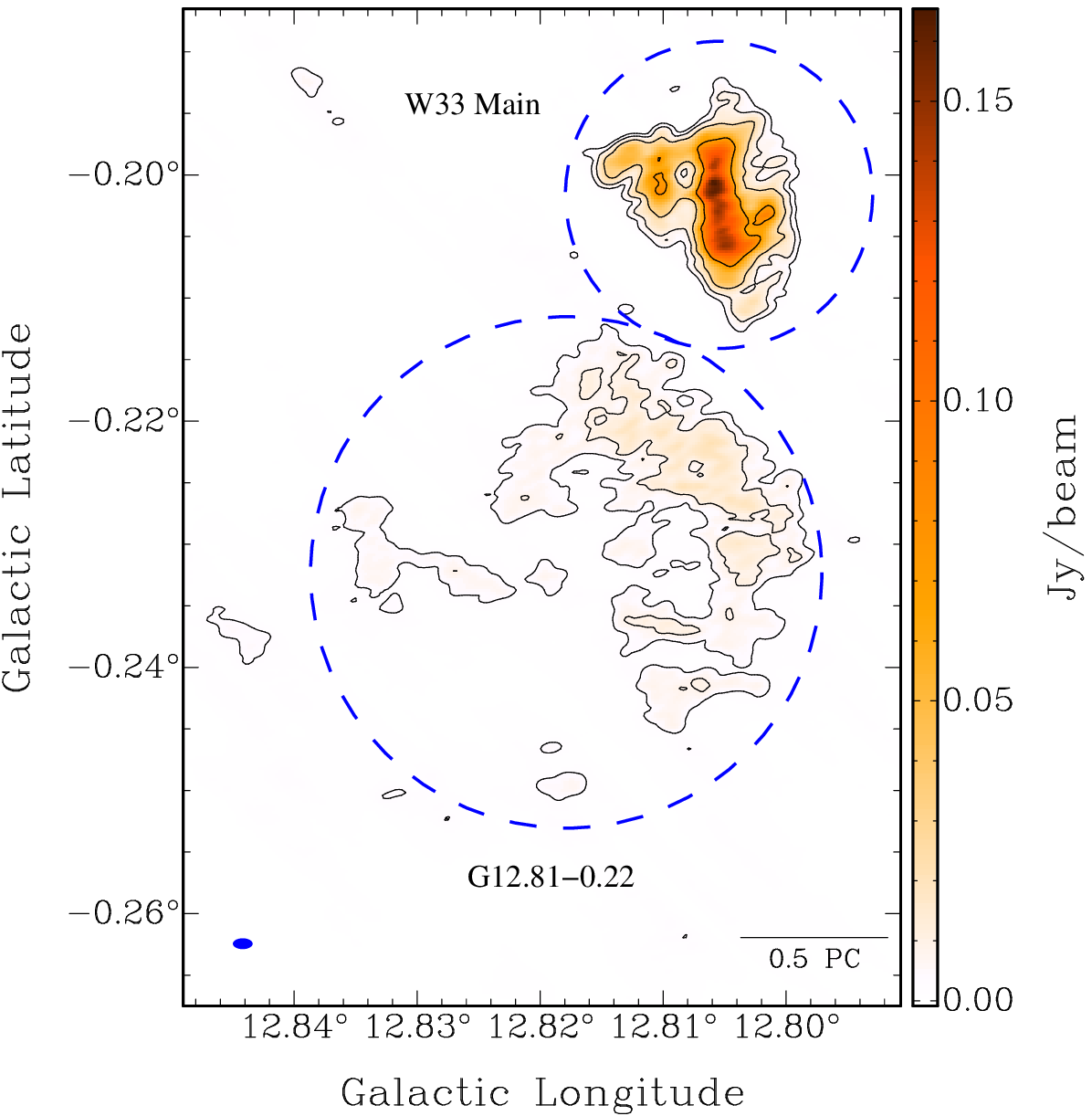}
                \includegraphics[scale=0.54]{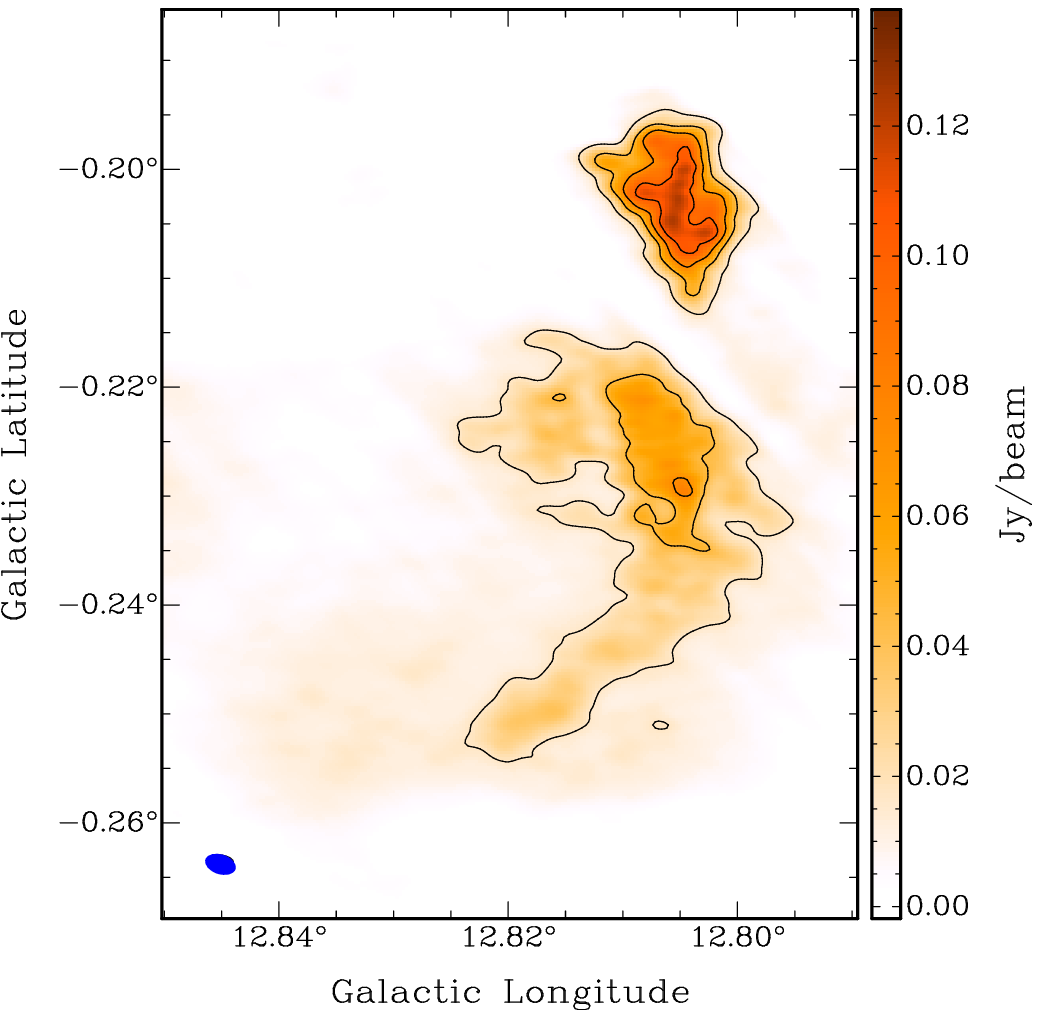}
                \includegraphics[scale=0.4]{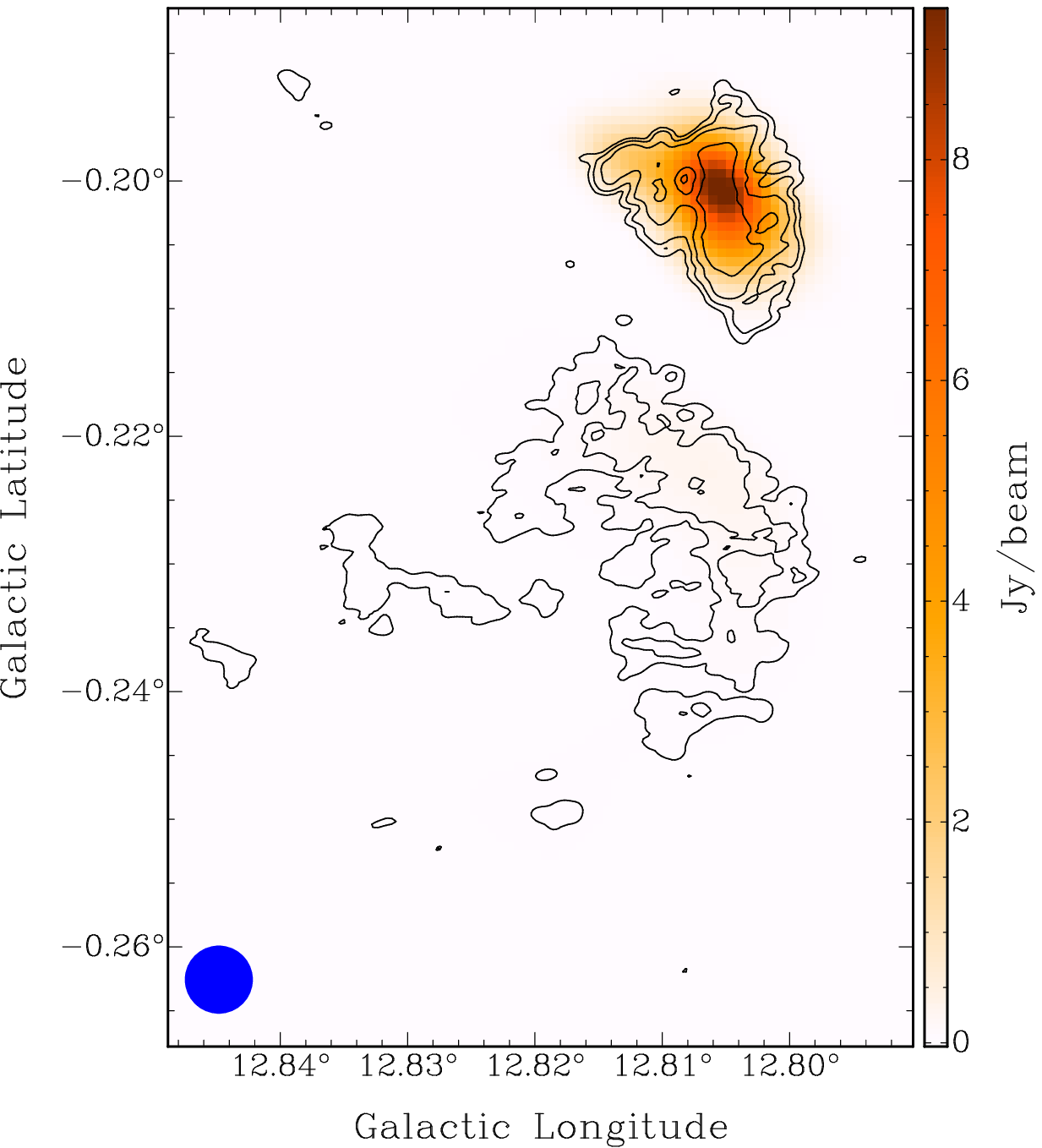}
                \caption{Radio continuum map of source W33 Main and G12.81$-$0.20 at 1400 MHz (left panel) and 610 MHz (center panel) after primary beam correction and rescaling. The contour levels in the 1400~MHz map are at 3$\sigma$ and factors of 3 thereafter. The contour levels in the 610~MHz map start at 3$\sigma$ with a linear step size of 5$\sigma$ thereafter, where $\sigma$ is rms noise in the maps equivalent to 0.91~mJy~beam$^{-1}$ and 5.6~mJy~beam$^{-1}$ at 1400 and 610~MHz, respectively. Right panel: GLOSTAR wide-band reference frequency (5.8~GHz) map overlaid with the GMRT 1400~MHz contour. The blue-filled ellipse shows the corresponding beam size.}
                \label{fig:cont_map}
        \end{figure*}

        The left and middle panels of Figure~\ref{fig:cont_map} shows the radio continuum emission toward the W33 Main region at 1400~MHz and 600~MHz.
    The angular size of the radio emission region of W33 Main is 64.8$\arcsec$ $\times$ 55.4$\arcsec$ which corresponds to linear sizes of 0.75~pc $\times$ 0.63~pc, respectively, at a distance of 2.4~kpc. Along with W33 Main, we also detected an arc-shaped diffuse continuum source, G12.81$-$0.22, which is located in southeast direction from W33 Main.
  The detail of radio continuum emission from both the sources at the GMRT frequencies are listed in Table~\ref{tab:emission_summ}.
        
        The radio continuum emission at 5.8~GHz with a resolution of 18$\arcsec$ as seen in the GLOSTAR survey is shown in Figure~\ref{fig:cont_map}. While there is good correspondence between the radio emission at 5.8 and 1.4~GHz, the latter shows several diffuse emission components which are not visible in the GLOSTAR VLA-D configuration continuum map. At the GLOSTAR reference frequency (5.8~GHz), details on radio continuum emission are listed in Table~\ref{tab:emission_summ}.

        The GMRT is an excellent instrument to look for and study the extended emission        around ultracompact HII regions given the high sensitivity to extended emission, with 7$\arcmin$ being the largest detectable structure at 1400~MHz which is equivalent to 4.9~pc at 2.4~kpc. In order to check for the presence of extended emission, we generated a radio continuum map of the region at a resolution of 40$\arcsec$ and determined the total radio flux. Other than the presence of the diffuse region G12.81$-$0.22 to the southwest of W33 Main, we did not find any evidence for extended emission around W33 Main. 
        
        \subsection{Spectral index map}\label{sec:spec_index}
        In order to investigate the nature of the radio continuum emission (thermal versus nonthermal), we examined the spectral index of the radio emission. A spectral index that is larger than $-$0.1 typically indicates thermal emission, while a value lower than $-$0.5 suggests that the emission is nonthermal. First, we constructed a spectral index map from the GMRT 610 and 1400~MHz observations. One of the problems encountered when creating spectral index maps from interferometer observations is that the uv coverage of the antennas is different for different frequencies, resulting in varying sensitivities in different angular scales. To minimize this effect, we created maps of the region at both frequencies using the same uv coverage. The uv coverage of the 610~MHz observation ranges from 0.13~k$\lambda$ to 50~k$\lambda$, while that of the 1400~MHz observation ranges from 0.21~k$\lambda$ to 121~k$\lambda$. We imaged the region using a common cell size, angular resolution (specified by the \texttt{BMAJ} and \texttt{BMIN} keywords), and a common uv range of 0.21~k$\lambda$ to 50~k$\lambda$. After this process, the beam size of both maps were 8.54$\arcsec$ $\times$ 4.15$\arcsec$ and the plate scale was 1$\arcsec$ per pixel.
        
\begin{figure*}[h!]%
                \centering
                \includegraphics[scale=0.4]{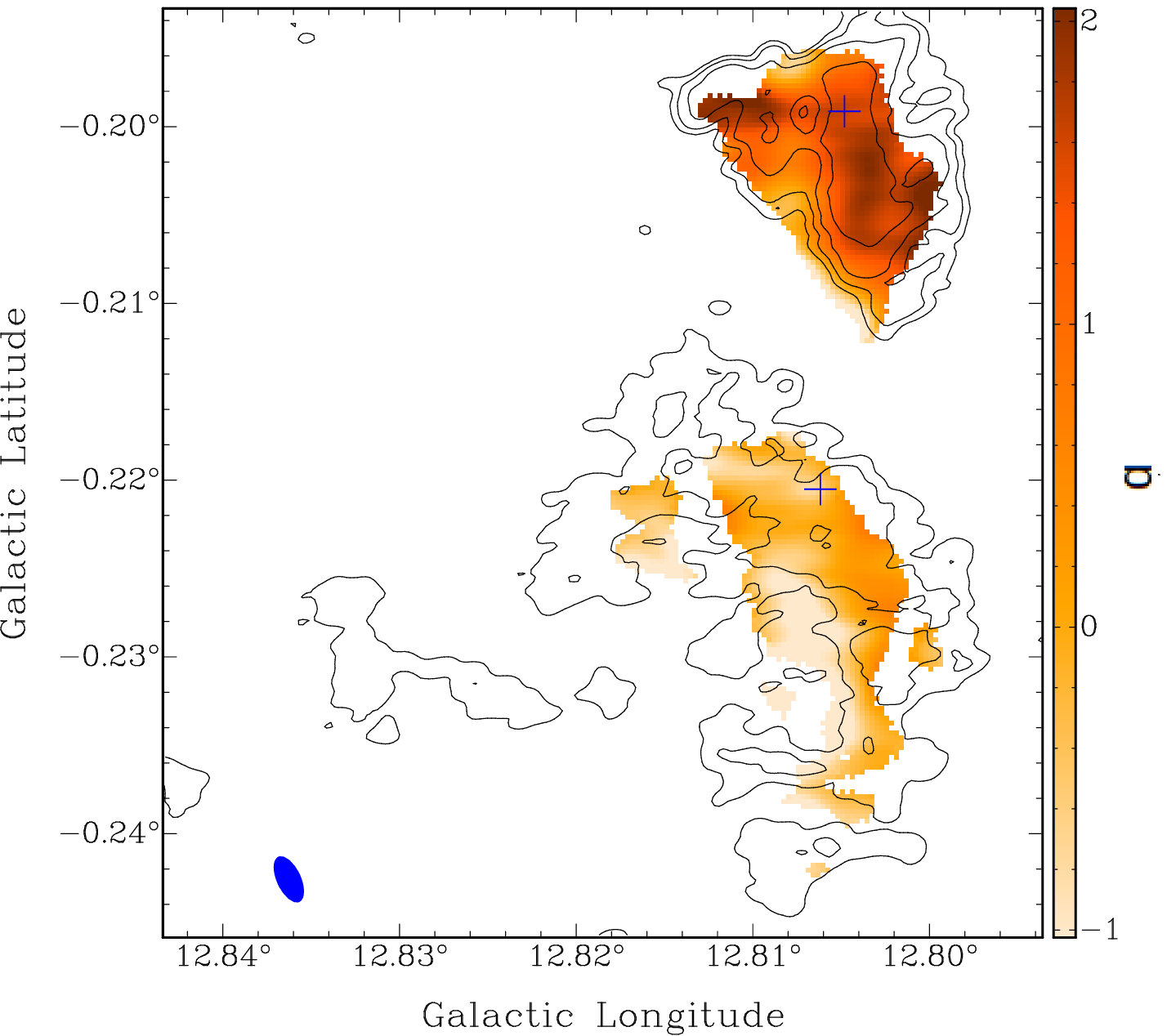}
                \includegraphics[scale=0.4]{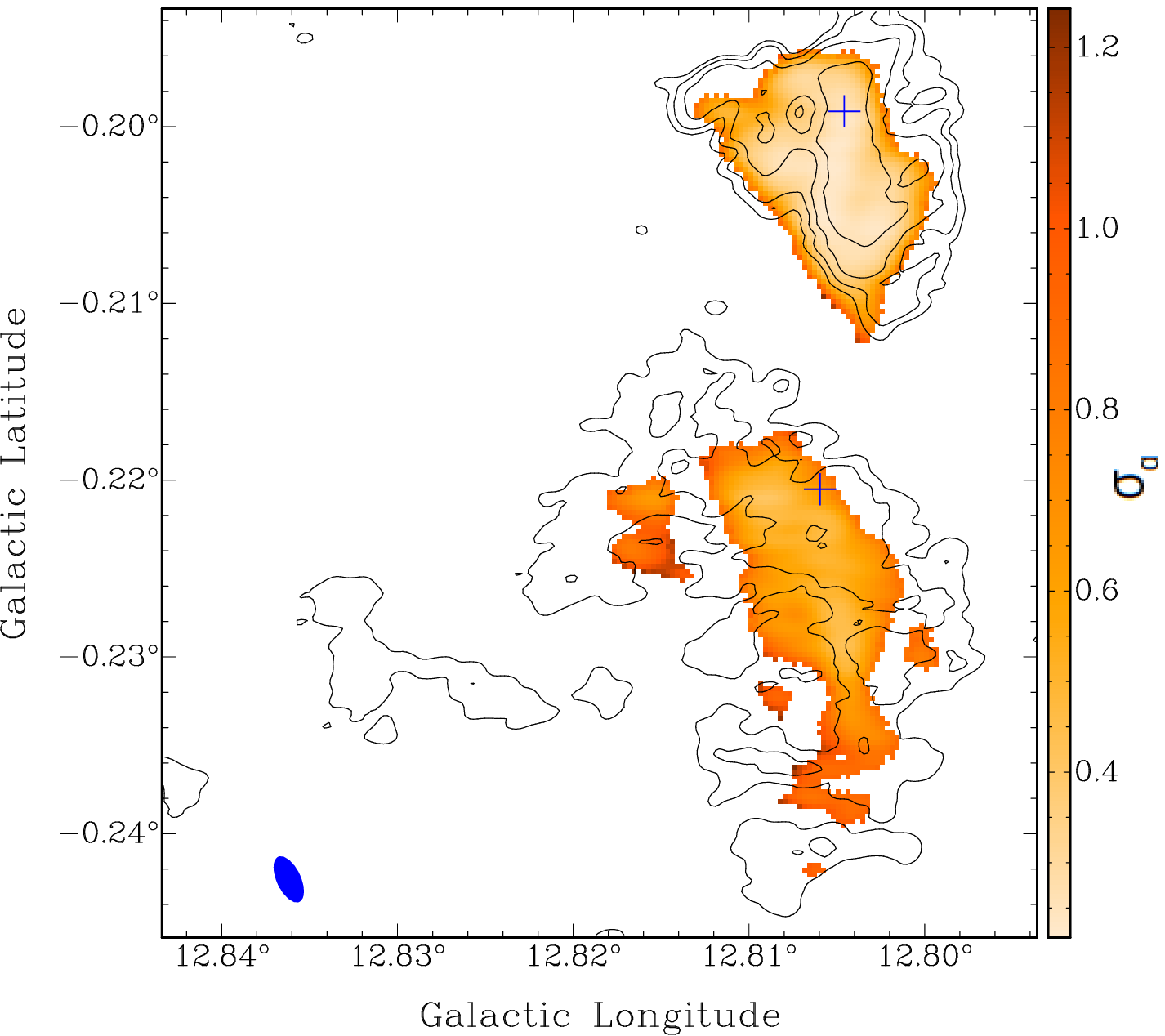}
                \caption{Maps of the spectral index between the continuum emission at 610~MHz and 1400~MHz (left panel) and its uncertainty (right panel) of W33 Main and G12.81$-$0.22, overlaid with contours of the radio emission at 1400~MHz at full resolution. The contour levels are as described in Figure \ref{fig:cont_map} (left panel). The blue-filled ellipse shows the corresponding beam size. The blue cross corresponds to peak emission toward W33 Main and G12.81$-$0.22 at 1400~MHz.}
                \label{fig:spex_map}
        \end{figure*}

        Since the process of self-calibration results in the loss of absolute astrometry, determining the spectral index from the maps above could lead to systematic errors due to a position offset between the two maps. To correct for the differential astrometric errors, we determined the positions of point sources that are detected at both frequencies using the task \texttt{JMFIT} and determined the position offset between the two maps. The 1400~MHz map was then shifted using the task \texttt{GEOM}, after which it was confirmed that the point sources line up on both maps accurately. Finally, the spectral index map was created using the task \texttt{COMB} with \texttt{OPCODE} set to "SPIX" for pixels having a radio continuum above the $5\sigma$ level at both frequencies. The spectral index map determined from this procedure is shown in the left panel of Figure~\ref{fig:spex_map}, with the uncertainty being shown in the right panel. 
        
\begin{figure*}[h!]%
                \centering
                \includegraphics[scale=0.4]{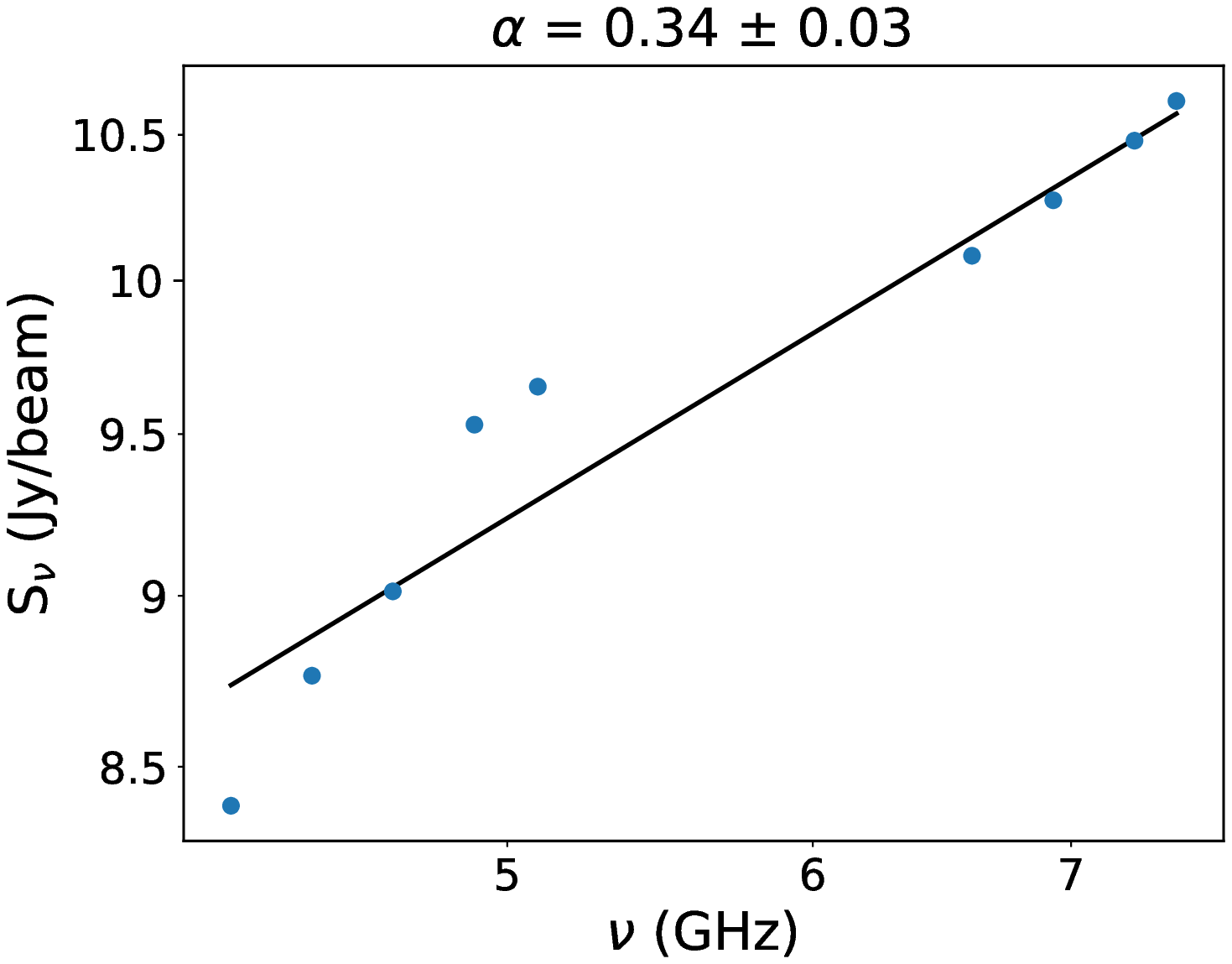}
                \includegraphics[scale=0.4]{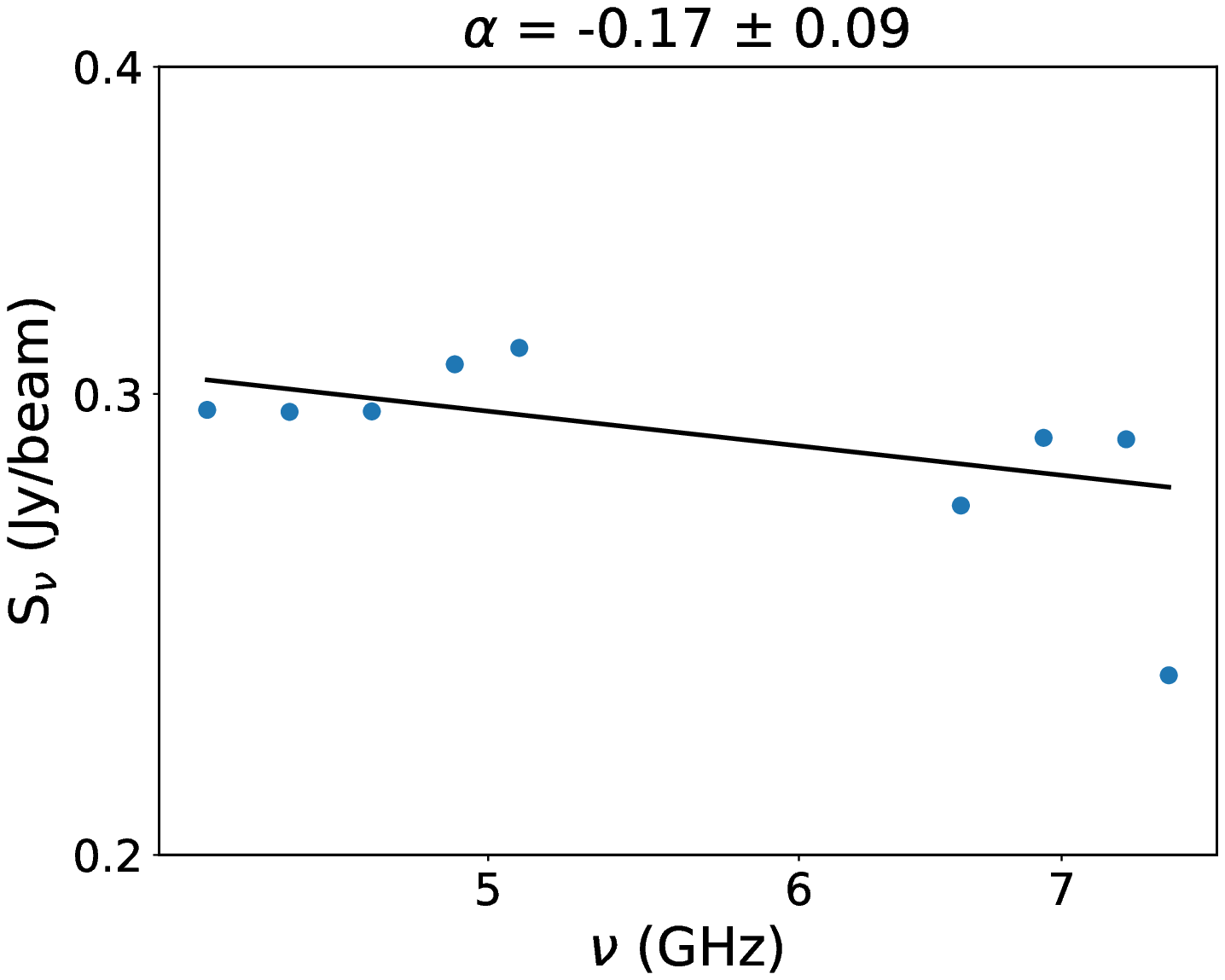}
                \caption{Spectral index over 4$-$8~GHz from the GLOSTAR survey corresponding to peak emission toward W33 Main (left panel) and G12.81$-$0.22 (right panel; see also Medina et al., in prep).}
                \label{fig:spec_fit}
        \end{figure*}

        Since the GLOSTAR survey provides radio continuum images at several frequency bins, we determined the spectral index by fitting the flux density as a function of frequency at each pixel. Figure~\ref{fig:spec_fit} shows the fit toward the peak emission of W33 Main and G12.81$-$0.22, and the values for the spectral index are tabulated in Table~\ref{tab:spec_index}.
        
        The spectral index determined from both GMRT observations and the GLOSTAR survey suggest that the emission is thermal in both sources. Furthermore, at the GMRT wavelengths, the continuum emission in W33 Main has a spectral index close to 2, while that in G12.81$-$0.22 is close to $-0.1$. At the location of peak emission, the spectral index between 610~MHz and 1.4~GHz is found to be $1.6 \pm 0.3$ and $-0.02 \pm 0.5$ for W33 Main and G12.81$-$0.22, respectively (Table~\ref{tab:spec_index}). At the frequencies corresponding to the GLOSTAR survey, the spectral index is found to be $0.35 \pm 0.04$ and $-0.17 \pm 0.09$ at the location of peak emission of W33 Main and G12.81$-$0.22, respectively (Medina et al., in prep). The W33 Main region shows a positive spectral index over the whole frequency range, while G12.81$-$0.22 has a negative spectral index. This shows that the optical depth of radio continuum emission from W33 Main is very high at 1.4~GHz, reducing to moderate values in the 4$-$8~GHz band, while the radio continuum from G12.81$-$0.22 is optically thin even at 610~MHz.
        
        \begin{table}[h!]
                \caption{Spectral index ($\alpha$) of peak emission toward W33 Main and G12.81$-$0.22 between different frequency ranges.}
                \label{tab:spec_index}
                \centering
                \begin{tabular}{c c c }
                        \hline \hline
                        Region & 0.61$-$1.4~GHz & 4.2$-$7.5~GHz \\
                        \hline
                        W33 Main & $1.6 \pm 0.3$ & $0.35 \pm 0.04$ \\
                        G12.81$-$0.22 & $-0.02 \pm 0.5$ & $-0.17 \pm 0.09$ \\
                        \hline
                \end{tabular}
        \end{table}
        
        \subsection{Radio recombination lines}\label{sec:rrl}
\begin{figure*}[ht!]
                \centering
                \includegraphics[scale=0.4]{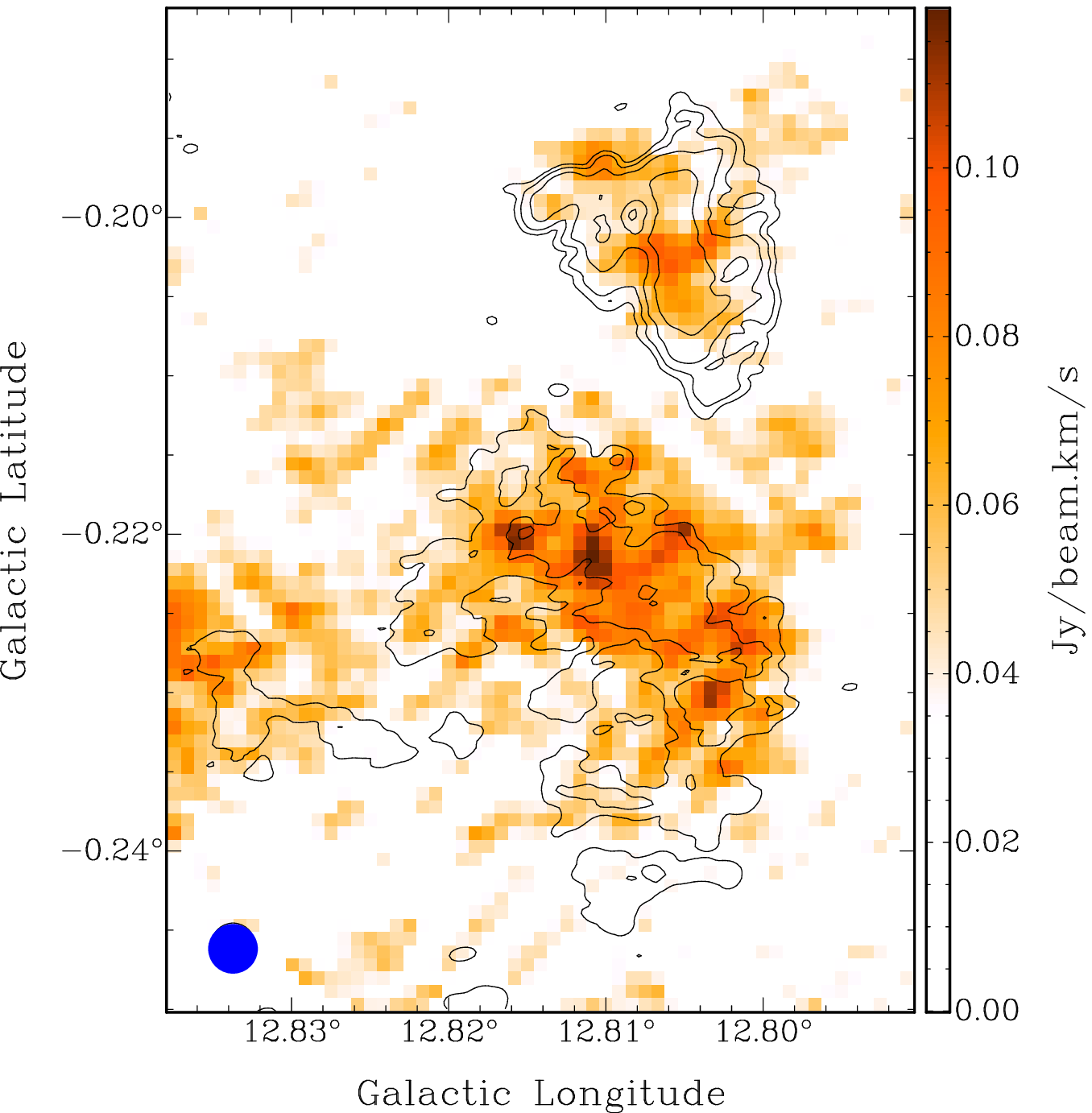}
                \includegraphics[scale=0.395]{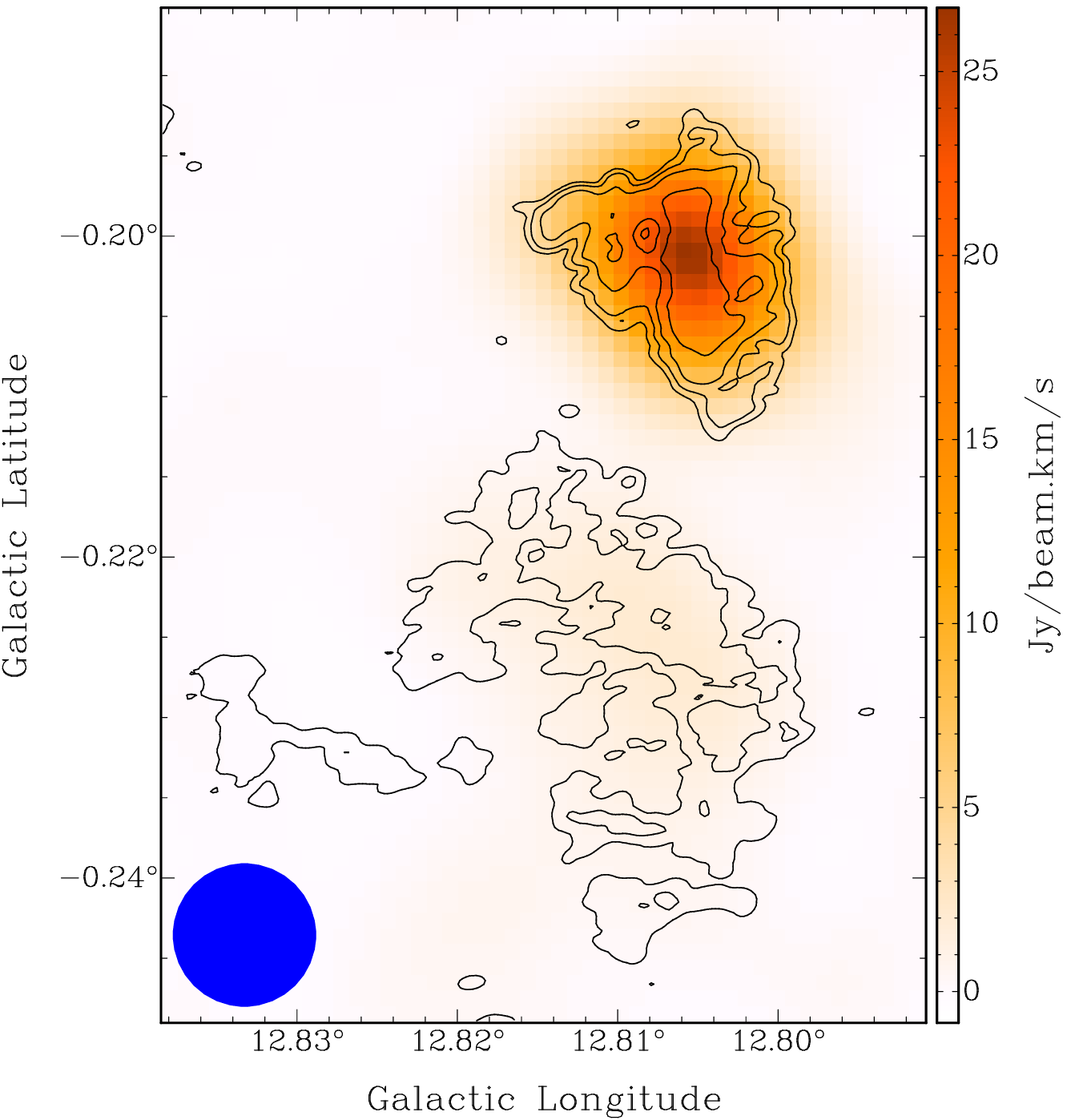}
                \caption{Integrated line intensity map toward W33 Main and G12.81$-$0.22. Left panel: GMRT H167$\alpha$ at 10$\arcsec$ resolution. Right panel: GLOSTAR stacked recombination line at 25$\arcsec$ resolution, overlaid with contours of the radio emission at 1400~MHz at full resolution. The contour levels are as described in Figure \ref{fig:cont_map} (left panel). The blue-filled circle shows the corresponding beam size.}  
                \label{fig:moment0}
        \end{figure*}

        The GMRT observations included the low-frequency H167$\alpha$ recombination line, which can provide information on the diffuse ionized gas around W33 Main. The H167$\alpha$ line was marginally detected toward W33 Main and G12.81$-$0.22 at the 4.7$\sigma$ and 4.4$\sigma$ level, respectively. Since the emission was observed to be fragmented at the full resolution of the array, we restricted the resolution to 10$\arcsec$ for our analysis. The peak line strength toward W33 Main and G12.81$-$0.22 was 7.5 and 7.1~mJy~beam$^{-1}$, respectively, while the 1$\sigma$ noise was 1.6~mJy~beam$^{-1}$. Figure~\ref{fig:moment0} shows the integrated intensity map of the H167$\alpha$ line. As expected, the line and the continuum emission are well correlated. Due to the poor S/N, the line profile could only be fitted at a few pixel locations. We are hence unable to trace the velocity field of ionized gas using the H167$\alpha$ line. The average line width of the H167$\alpha$ line was found to be $19.6 \pm 7.5$ and $18.1 \pm 5.9$~km~s$^{-1}$ toward W33 Main and G12.81$-$0.22, respectively. The line width in W33 Main is smaller than what is typically expected in compact HII regions.
        
        The GLOSTAR survey provides recombination line data cubes by stacking six different recombination lines with a spectral resolution of 5~km~s$^{-1}$ and spatial resolution of 25$\arcsec$. The stacked recombination line was detected with a very good S/N and correlates well with the continuum emission at 1.4 and 5~GHz as shown in Figure~\ref{fig:moment0}. The peak line strength toward W33 Main and G12.81$-$0.22 was 680 and 83~mJy~beam$^{-1}$, respectively, while the 1$\sigma$ noise in a line-free channel is 4.0~mJy~beam$^{-1}$.  The average line width of the GLOSTAR stacked recombination line was found to be $36.3 \pm 1.1$ and $23.6 \pm 1.4$~km~s$^{-1}$ toward W33 Main and G12.81$-$0.22, respectively, which are comparable to typical line widths found in the compact and diffuse HII regions, respectively.

        \subsection{Electron temperature and photon hardening}\label{sec:electrontemp}
        The electron temperature can be determined using the radio continuum and recombination line data as long as the optical depth of the continuum emission is not very high. The methodology is described in Appendix~\ref{sec:dev_te}. As discussed in \S~\ref{sec:spec_index}, while the radio continuum is optically thick in W33 Main at 1.4~GHz, the optical depth is moderate at the frequencies covered by the GLOSTAR survey.
        \begin{figure}[h!]%
                \centering
                \includegraphics[scale=0.5]{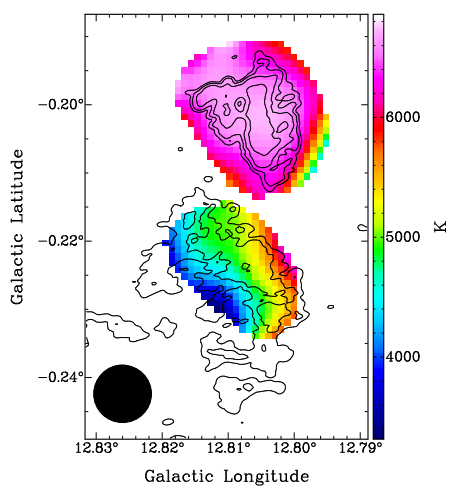}
                \caption{Electron temperature map of W33 Main and G12.81$-$0.20 overlaid with contours of the radio emission at 1400~MHz at full resolution. The contour levels are as described in the left panel of Figure~\ref{fig:cont_map}. The black-filled circle represents the GLOSTAR beam size.} \label{fig:ele_temp_glostar}
        \end{figure}
        
        Figure~\ref{fig:ele_temp_glostar} shows the map of electron temperature in the region determined from the GLOSTAR survey. The average electron temperature toward W33 Main is found to be $6343 \pm 222$~K. This is consistent with earlier studies. \citet{bienging} found an electron temperature of 6000~K, while \citet{quireza:62} found an estimate of $7620 \pm 100$~K, with both studies using the H91$\alpha$ line, but at different angular resolutions. We are unable to determine the electron temperature at 1.4~GHz using the GMRT data toward the central regions of W33 Main due to the high optical depth of continuum emission. At the periphery of the HII region, where the optical depth is moderate, the electron temperature is measured to be $9145 \pm 4710$~K, where the large uncertainty is due to the poor S/N of the GMRT line data. Although the electron temperature cannot be determined from the recombination line when the radio continuum is optically thick, one can use the latter to obtain an estimate of the electron temperature. This is because the specific intensity of radio continuum emission is equal to the blackbody function of the electron temperature when the optical depth is very high, that is to say $I_\nu = B_\nu(T_e)$. Using this approach, we obtain an electron temperature of 10,000 to 12,100~K toward the central region of W33 Main. This is comparable to the results of \citet{gardner..h134}, who derived the electron temperature toward peak emission to be between 10600~K to 12000~K using the H134$\alpha$ line. For the diffuse continuum source G12.81$-$0.22, the electron temperature is measured to be $4843 \pm 355$~K using GLOSTAR RRLs, and $6045 \pm 2590$~K using the GMRT observation of the H167$\alpha$ recombination line. 
        
        It can be seen that the electron temperature measured toward the central regions of W33 Main are significantly higher in GMRT compared to GLOSTAR. In contrast, the electron temperature measured toward G12.81$-$0.22 is broadly consistent at low and high frequencies. This is very likely due to the effects of photon hardening toward the boundary of HII regions \citep{2004MNRAS.353.1126W}. Since the bound-free absorption cross-section is lower at higher photon energy, the ionization at the edges of HII regions with a high optical depth is primarily driven by higher energy photons. Since the excess energy of the photons is converted into kinetic energy of the electrons, the electrons are expected to have a higher temperature. In the case of W33 Main, since the optical depth is high at 1.4~GHz, the continuum emission mainly originates from the outer edges of the HII region. Consequently, the electron temperature inferred from the radio continuum at 1.4~GHz is significantly higher than what is measured at GLOSTAR frequencies where the continuum optical depth is much lower. In G12.81$-$0.22, however, the optical depth of the radio continuum is low, even at 1.4~GHz, and the electron temperature measured using GMRT is found to be consistent with that of GLOSTAR. Even in G12.81$-$0.22, taking the center of the HII region to be the center of the arc, Figure~\ref{fig:ele_temp_glostar} shows that the electron temperature increases toward the outer regions, which may also be due to photon hardening. While noting that,~\citet{2015ApJ...812...45W} did not find evidence of photon hardening in Orion A and since the phenomenon of photon hardening is  primarily discussed in the context of the warm diffuse interstellar medium of a galaxy, our observations suggest that it is detectable within individual HII regions as well.

        \subsection{Gas dynamics}\label{sec:gasdynamics}
        \begin{figure}[h!]%
                \centering
                \resizebox{\hsize}{!}{\includegraphics[scale=0.4]{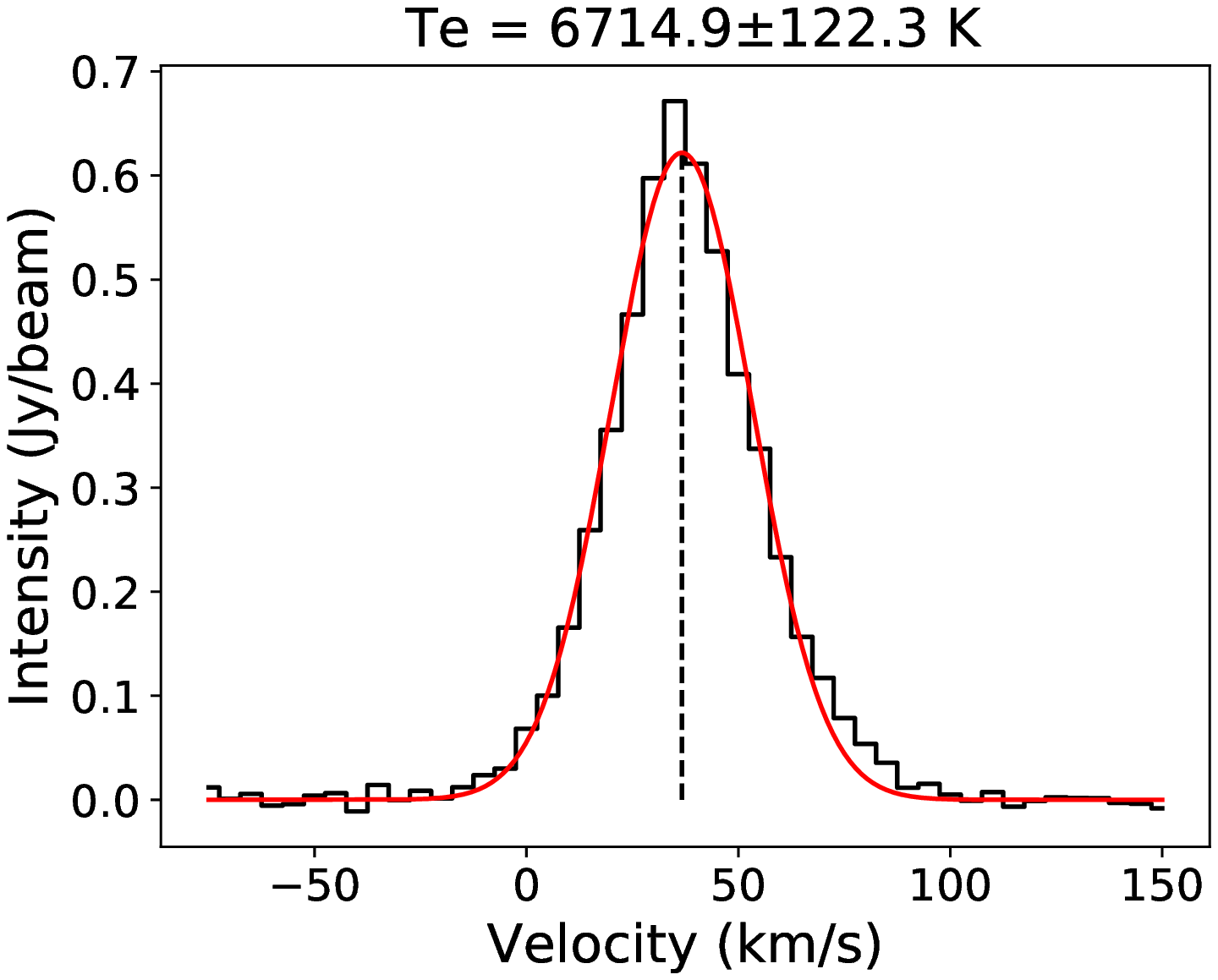}
                        \includegraphics[scale=0.4]{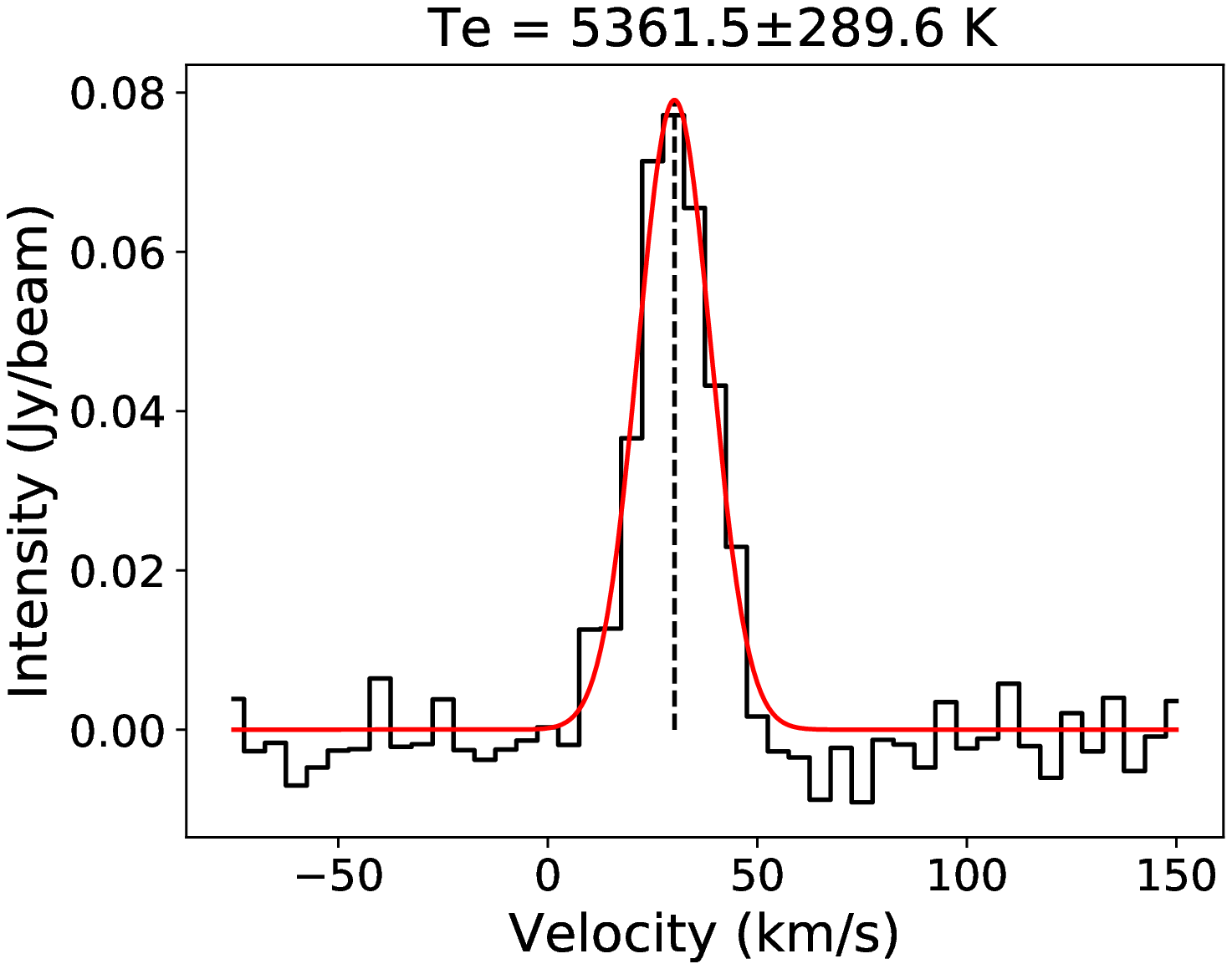}}
                \caption{Example of Gaussian fitted GLOSTAR stacked recombination line corresponding to the peak emission toward W33 Main (left panel) and G12.81$-$0.22 (right panel). Toward the peak emission of W33 Main, the line amplitude, line width, and velocity were found to be 622 $\pm$ 8.6~mJy/beam, 39.3 $\pm$ 0.6~km~s$^{-1}$, and 36.7 $\pm$ 0.3~km~s$^{-1}$, respectively. For G12.81$-$0.22, toward the peak emission, the line amplitude, line width, and velocity were found to be 79.1 $\pm$ 2.9~mJy/beam, 20.4 $\pm$ 0.9~km~s$^{-1}$ and 30.2 $\pm$ 0.4~km~s$^{-1}$, respectively.}\label{fig:rrl_fit}
        \end{figure}
        
\begin{figure*}[h!]%
                \centering
                \includegraphics[scale=0.45]{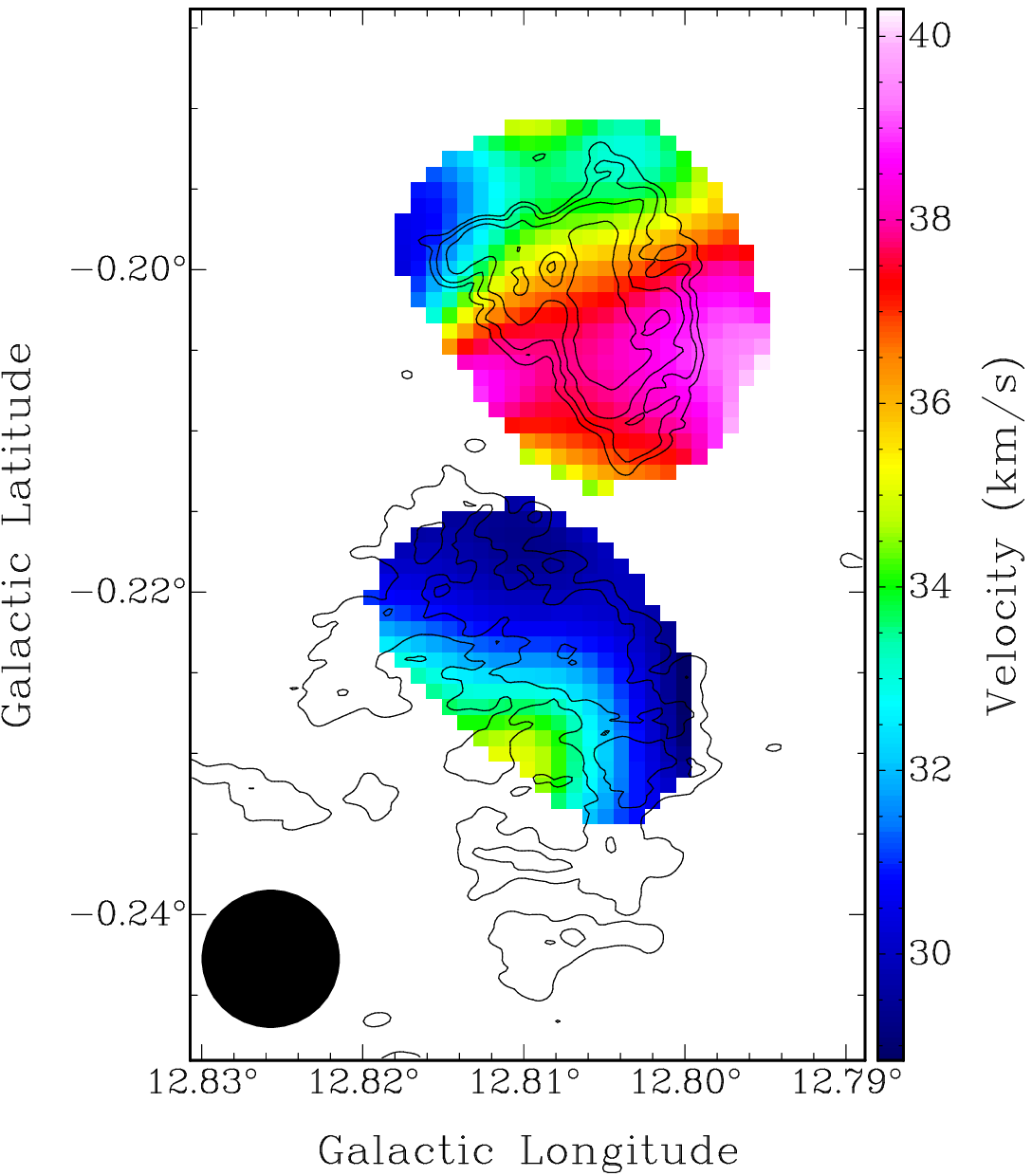}
                \includegraphics[scale=0.45]{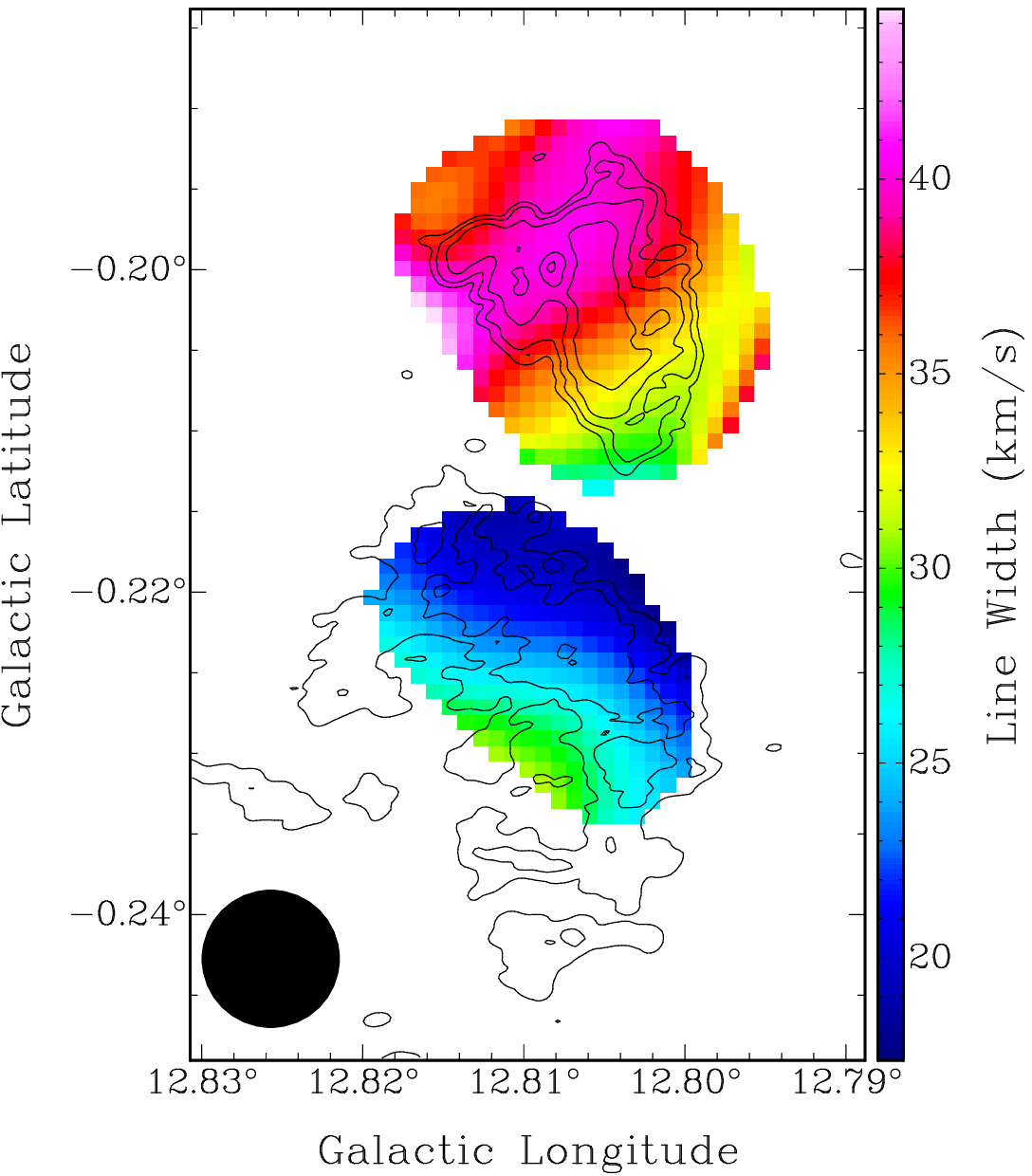}
                \caption{Velocity field and a map of the line width obtained from the Gaussian fitting of GLOSTAR stacked recombination line data is shown in the left and right panels respectively. The maps are overlaid with contours of the radio emission at 1400~MHz at full resolution. The contour levels are as described in Figure \ref{fig:cont_map} (left panel). The black-filled circle represents the GLOSTAR beam size.}\label{fig:maps}
        \end{figure*}

        To investigate the dynamics of the ionized gas, we performed Gaussian fits to the GLOSTAR stacked recombination line data on a per pixel basis, with Figure~\ref{fig:rrl_fit} showing sample fits toward the peak emission. The resulting maps of the velocity field and line width are shown in Figure~\ref{fig:maps}. The velocity field shows the presence of a velocity gradient in both the sources. In the case of W33 Main, the velocity increases from northeast to southwest with a velocity difference of around 10~km~s$^{-1}$. In the case of G12.81$-$0.22, there is a velocity gradient across the arc region in the radial direction, suggestive of expansion.
        The ratio of pressure in the HII region and the associated dust clump is given by the following: 
        \begin{equation}\label{eq:pressure}
        \frac{P_{HII}}{P_{Dust}} = \frac{2<n_e>T_e}{<n_D>T_D}
        ,\end{equation}
        where $<n_e>$ and $T_e$ are the mean electron density and electron temperature of the HII regions, respectively (Table~\ref{tab:physicalpara}); and $<n_D>$ and $T_D$ are the mean hydrogen density and dust temperature of the associated dust clump, respectively (see \S\ref{sec:cold_dust}). For the HII regions toward W33 Main (regions 1, 2, and 3 as mentioned in \S\ref{sec:physical}), the ratio is 19, 13, and 20, respectively, which is $\gg$ 1. This suggests the ongoing pressure-driven expansion of HII regions. However, for G12.81$-$0.22, the ratio is close to unity, which indicates that the pressure-driven expansion for this HII region is coming to a halt.
        
        Another interesting feature seen in the right panel of Figure~\ref{fig:maps} is the presence of large line widths toward the northern region of W33 Main. W33 Main has an average electron temperature of 6360 K, which corresponds to a thermal line width of 17~km~s$^{-1}$. The northern region of W33 Main has a line width that is greater than 40~km~s$^{-1}$, which is significantly broader than the thermal broadening of the line. Following \citet{keto_2008rrl}, the recombination lines have a Voigt profile with a line width $\Delta\nu_V$ given by
        \begin{equation}\label{eq:rrllw1}
        \Delta\nu_V = 0.5343\Delta\nu_I + \left[\Delta\nu_G^2 + (0.4657\Delta\nu_I)^2 \right]^{1/2}
        ,\end{equation}
        where $\Delta\nu_I$ is the Lorentzian line width due to pressure broadening and $\Delta\nu_G$ is the combination of thermal and dynamical broadening, given by
        \begin{equation}\label{eq:rrllw2}
        \Delta\nu_G = \sqrt{\Delta\nu_t^2 + \Delta\nu_d^2}
        .\end{equation}
        Here, $\Delta\nu_t$ and $\Delta\nu_d$ refer to the thermal and dynamical broadening, respectively. 
        The ratio of pressure and thermal broadening is given by~\citep{1972MNRAS.157..179B},
        \begin{equation}
        \frac{\Delta\nu_I}{\Delta\nu_t} = 0.142\left( \frac{n}{100} \right)^{7.4} \left( \frac{n_e}{10^4} \right) 
        ,\end{equation}
        where $n$ is transition number and $n_e$ is the electron density in cm$^{-3}$. For the GLOSTAR survey, six recombination lines are observed, with the largest value of $n$ being 114. The northern region of W33 Main has an electron density of 6.2 $\times$ 10$^3$~cm$^{-3}$ (see Table~\ref{tab:physicalpara}), for which $\Delta\nu_I/\Delta\nu_t \approx 0.23$. Thus, the line width from pressure broadening is around 3.9~km~s$^{-1}$. Thus, from eqs.~(\ref{eq:rrllw1}) and (\ref{eq:rrllw2}), a line width in excess of 40~km~s$^{-1}$ implies dynamical broadening in excess of 30~km~s$^{-1}$. \citet{2021ApJ...921..176B} investigated the kinematic properties of Galactic HII regions using RRL emission and suggest that the HII regions with a center-peaked RRL width and velocity gradient favor solid-body rotation. In the case of W33 Main, by assuming the broad line width region to be the center, the velocity gradient structure suggests rotation motion. \citet{2021ApJ...921..176B} also report a velocity gradient for W33 Main. However, the spatial resolution of the GLOSTAR RRL map is increased compared to the observations used by \citet{2021ApJ...921..176B}. Because of this, the velocity gradient does not appear to be homogeneous, but has a complex velocity structure. W33 Main consists of three HII regions (\citet{Anderson_2014}; see also Figure~\ref{fig:region} and \S\ref{sec:physical}), which have slightly different average velocities (Figure~\ref{fig:maps}). Also, the broad radio recombination lines are located toward the boundary of these HII regions. This indicates some complicated dynamic motions of the ionized gas inside W33 Main. For a detailed examination of the kinematics of the region, high-resolution molecular line observations are required.

        \subsection{Physical properties of the ionized gas}\label{sec:physical}
\begin{figure}[h!]
                \centering
                \includegraphics[scale=0.6]{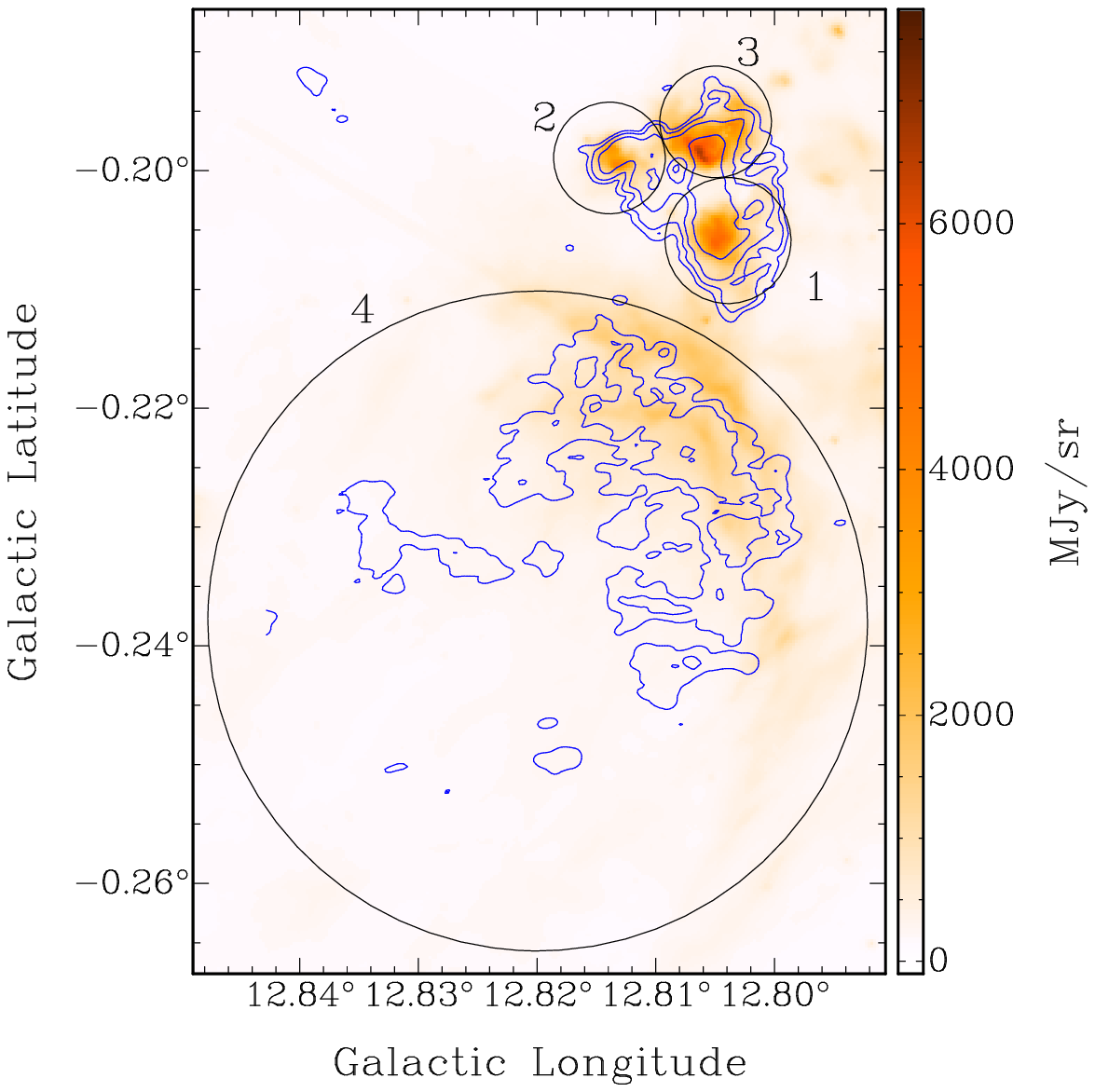}
                \caption{GLIMPSE 8~$\mu$m map overlaid with blue contours of the radio emission at 1400~MHz at full resolution. The contour levels are as described in the left panel of Figure \ref{fig:cont_map}. The black circles represent the HII regions toward W33 Main and G12.81$-$0.22 taken from the WISE catalog of Galactic HII regions~\citep{Anderson_2014}. Some regions have been slightly adjusted in order to avoid overlap and for better correlation with radio and 8~$\mu$m emission.\label{fig:region}}
        \end{figure}

        \citet{Anderson_2014} made a catalog of over 8000 HII regions using data from the WISE satellite based on the morphology of mid-infrared emission. In the W33 Main region, four HII regions were identified (Table~\ref{tab:wise}), of which three are toward W33 Main and one is toward the diffuse continuum source, G12.81$-$0.22 (see Figure~\ref{fig:region}).
        
        \begin{table*}[h!]
                \caption{HII regions in the W33 Main region from the WISE catalog of Galactic HII regions of~\cite{Anderson_2014}.}
                \label{tab:wise}
                \centering
                \begin{tabular}{c c c c c c c}
                        \hline \hline
                        Region & WISE & RA (J2000) & Dec (J2000) & Radius & Radius (used) &Velocity\\
                        & & h:m:s & d:m:s & $\arcsec$ & $\arcsec$ & km s$^{-1}$ \\
                        \hline
                        1 & G12.804$-$00.207 & 18:14:14.89 & $-$17:55:57.3 & 21 & 19 & 35.8 \\
                        2 & G12.813$-$00.200 & 18:14:14.55 & $-$17:55:13.6 & 21 & 17 & 35.8 \\
                        3 & G12.805$-$00.196 & 18:14:12.80 & $-$17:55:36.9 & 21 & 17 & 35.8 \\
                        4 & G12.820$-$00.238 & 18:14:23.91 & $-$17:56:01.8 & 126& 100 & $\ldots$ \\
                        \hline
                \end{tabular}
        \end{table*}
        
        One can determine the physical properties of the HII regions using the electron temperature (Fig.~\ref{fig:ele_temp_glostar}) and the distance to the source; the results are tabulated in Table~\ref{tab:physicalpara}. The radii of the different regions are slightly different from that given by \citet{Anderson_2014} to avoid overlap between the different regions and for better correlation between the infrared and radio emission. The peak emission measure can be estimated from the optical depth and electron temperature determined at the peak continuum emission using eq.~(\ref{eq:tau_c_n}). The average electron density, Lyman continuum photon rate, and ionized mass are estimated using the equations given in Appendix~\ref{sec:phy_pro}. Assuming that all the ionizing radiation comes from a single star, one can estimate the spectral type of the ionizing star using the Lyman continuum photon rate from \citet{2005A&A...436.1049M}. Table~\ref{tab:physicalpara} shows that the properties of W33 Main are consistent with it being an ultracompact HII region.

        \begin{table*}
                \caption{Physical properties of regions.}
                \label{tab:physicalpara}
                \centering
                \begin{tabular}{c c c c c c c c c}
                        \hline \hline
                        Region & Radius & $T_e$ & $\tau_C $ & EM & $n_e$ & $N_{\text{Ly}}$ & Ionized mass & Spectral type\\
                        & pc & K & (5.8~GHz) & $10^6$ pc cm$^{-6}$ & $10^3$ cm$^{-3}$ & $10^{48}$ s$^{-1}$ & $M_\sun$ &     \\
                        \hline
                        1 & 0.12 & $6522 \pm 158$ & $0.18 \pm 0.005$ & $12.4 \pm 0.1$  & $9.4 \pm 1.0$   & $5.4 \pm 1.1$ & $0.91 \pm 0.1$ & O6.5 \\
                        2 & 0.08 & $6528 \pm 239$ & $0.07 \pm 0.003$ & $5.02 \pm 0.06$ & $6.2 \pm 0.6$   & $1.7 \pm 0.3$ & $0.43 \pm 0.04$ & O8  \\
                        3 & 0.08 & $6487 \pm 151$ & $0.19 \pm 0.005$ & $13.0 \pm 0.1$  & $10.0 \pm 1.0$  & $4.3 \pm 1.0$ & $0.69 \pm 0.07$ & O7 \\
                        4 & 0.25 & $4856 \pm 356$ & $0.01 $            & $0.33 \pm 0.01$ & $0.38 \pm 0.04$ & $1.7 \pm 0.3$ & $6.9 \pm 0.7$   & O8 \\
                        \hline
                \end{tabular}
                \tablefoot{The columns show the radius, average electron temperature, peak continuum optical depth at 5.8~GHz, emission measure, electron density, Lyman continuum photon rate, ionized mass, and the spectral type of the central star for each of the four HII regions in the W33 Main region.}
        \end{table*}

        \subsection{Emission from cold dust}\label{sec:cold_dust}
\begin{figure*}[h!]%
                \centering
                \includegraphics[scale=0.4]{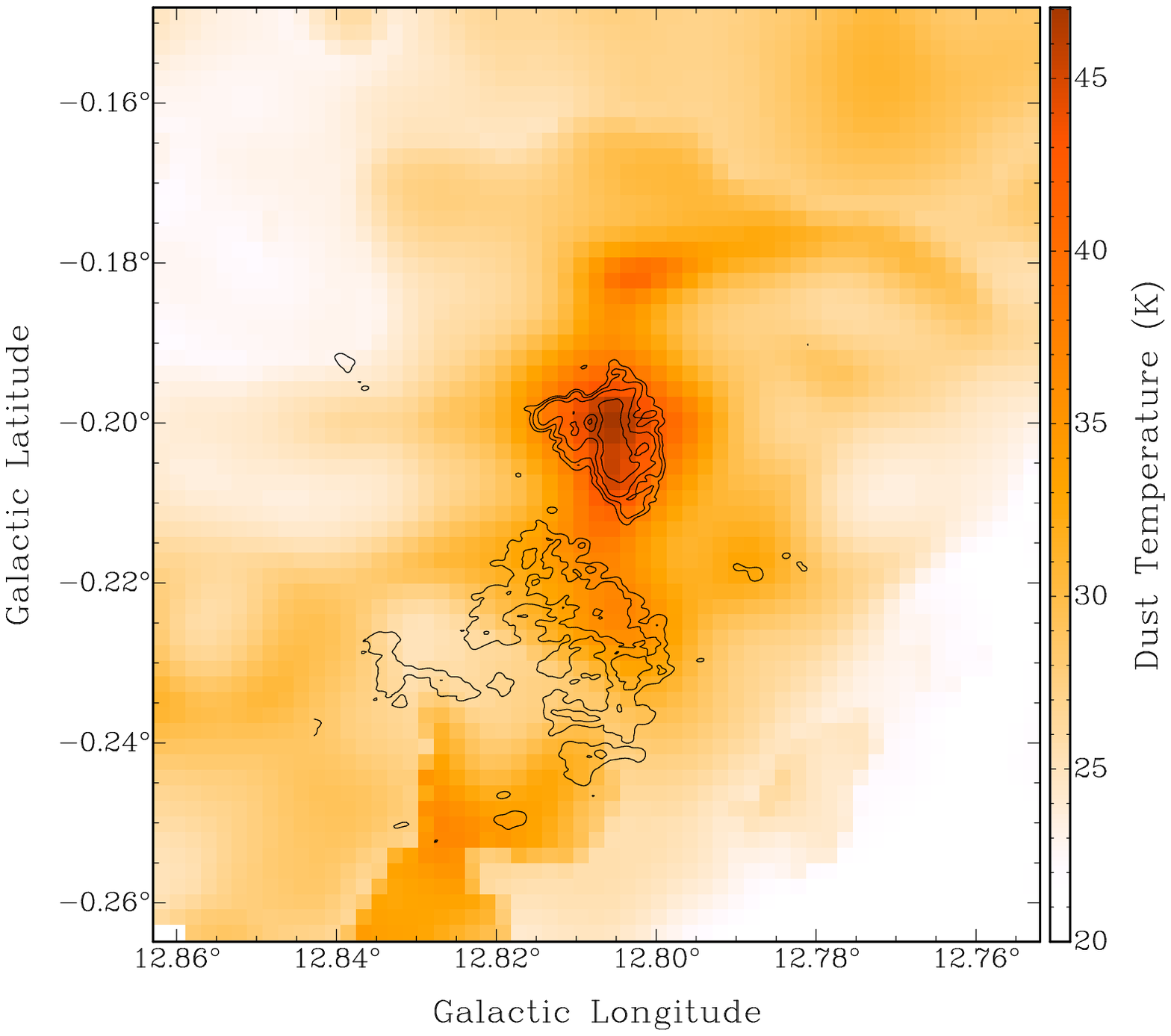}
                \includegraphics[scale=0.4]{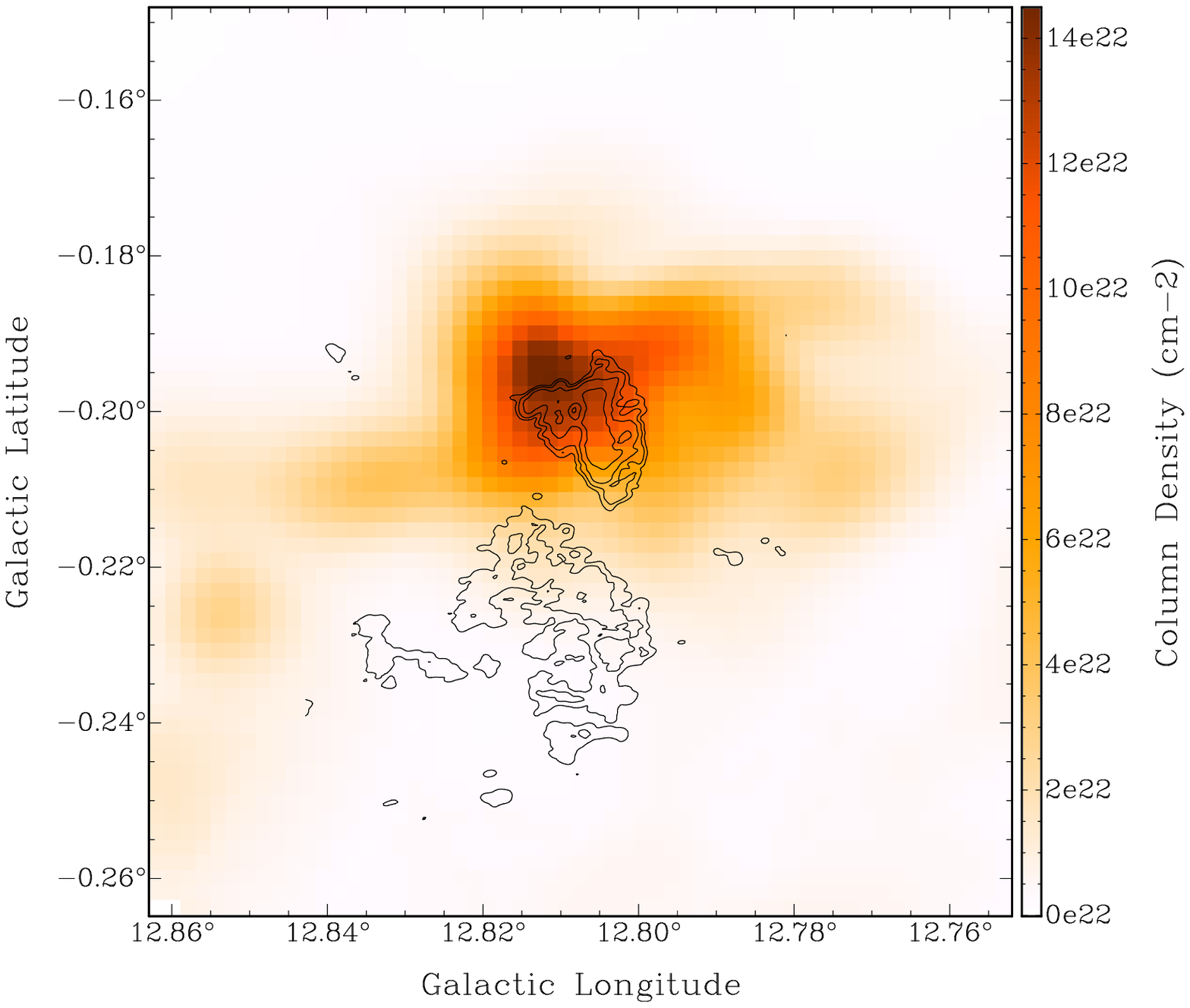}
                \caption{Maps of the dust temperature (left panel) and H$_2$ column density (right panel). The 1400~MHz radio continuum is overlaid as black contours with the contour levels as described in Figure~\ref{fig:cont_map} (left panel).}\label{fig:cold_dust}
        \end{figure*}

        Figure~\ref{fig:cold_dust} shows the map of the dust temperature determined from the pixel-by-pixel fitting of the dust SED from 870 to 70~$\mu$m. We derived the column density of molecular hydrogen $N$(H$_2$) from the dust optical depth at 500~$\mu$m using the following equation:
        \begin{equation}
        \tau_\nu = \mu m_H\kappa_\nu N(H_2)/R
        ,\end{equation}
        where $m_H$ is the mass of the hydrogen atom, $\mu$ is  the mean molecular weight, $\kappa_\nu$ is the dust opacity at 500~$\mu$m, and $R$ is the gas-to-dust ratio. We have taken $\mu$ to be 2.8, assuming a hydrogen
mass fraction of 0.7 (\citealt{kauffmann:5}); the dust opacity at 500~$\mu$m to be 5.04~cm$^2$~g$^{-1}$, which is appropriate for dust grains with thin ice mantles \citep{ossenkopf:8}; and $R$ to be 100. The right panel of Figure~\ref{fig:cold_dust} shows the distribution of the column density of molecular hydrogen in the region.
        
        The peak dust temperature is seen to be $\sim 45$~K in the region. This is consistent with the dust temperature derived by \citet{2014A&A...572A..63I} who found a value of $42.5 \pm 12.6$~K. The dust temperature in the region is significantly higher than the mean kinetic temperature (based on NH$_3$ observations) toward infrared dark clouds (14.8~K; \citealt{pillai:13}) and 6.7~GHz methanol maser sites (26.0~K; \citealt{pandian:14}). The high dust temperature can be inferred to reflect the more evolved state of the region. The peak column density of the clump in W33 Main is found to be $1.47 \times 10^{23}$~cm$^{-2}$ and the average column density in the region is $5.0 \times  10^{22}$~cm$^{-2}$. \citet{2014A&A...572A..63I} found the peak H$_2$ column density ($4.6 \pm 1.6 \times 10^{23}$~cm$^{-2}$), which is higher than what is found in our study. However, this is in part due to them using the dust opacity from \citet{hildebrand:6}, which is a factor of 2 smaller than that of \citet{ossenkopf:8} which have used in our study.
        
        The total mass of the W33 Main molecular clump is estimated to be $4.6 \times 10^{3}$~M$_\sun$. This is consistent with the observation of \citet{2014A&A...572A..63I} who estimated the mass to be $4.0 \pm 2.5 \times 10^{3}$~M$_\sun$. The mean hydrogen number density is found to be $4.7 \times 10^4$~cm$^{-3}$, assuming an effective radius of 0.7~pc. The effective radius is obtained by Gaussian fitting of the clump and is given by r$_{eff}$ $=$ $(\theta_1 \theta_2/\pi)^{1/2}$, where $\theta_1$ and $\theta_2$ are the FWHM of the major and minor axes of the Gaussian, respectively. \citet{2018PASJ...70S..50K} used the C$^{18}$O $J=1-0$ observation to find the mass, number density, and mean hydrogen column density of W33 Main to be $9.5 \times 10^3$~M$_\sun$, $1.5 \times 10^5$~cm$^{-3}$, and $5.1 \times 10^{22}$~cm$^{-2}$, respectively, which is roughly consistent with our value.

        \subsection{Young stellar object population}
        
\begin{figure*}[h!]
                \centering
                \includegraphics[scale=0.4]{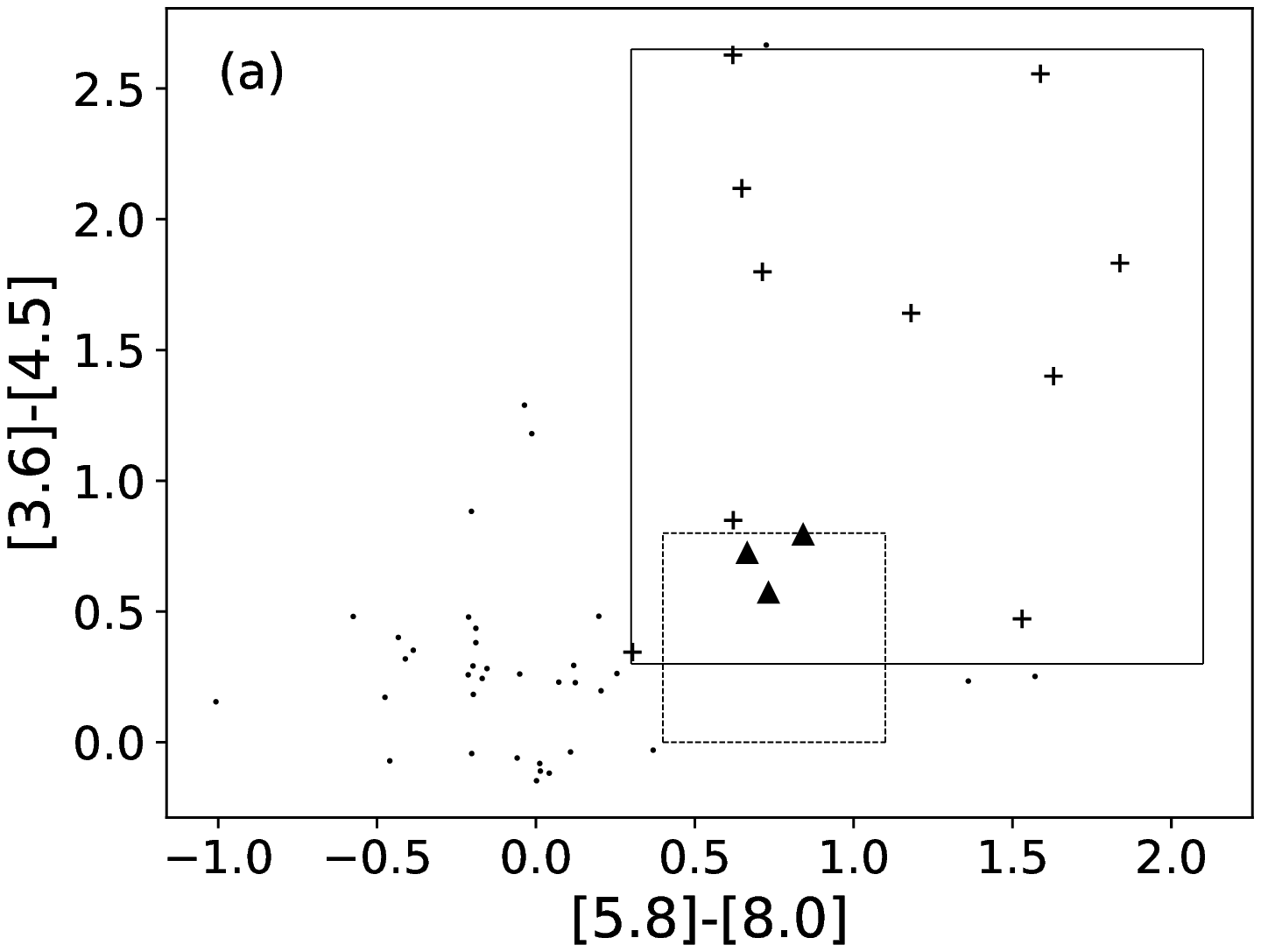}
                \includegraphics[scale=0.4]{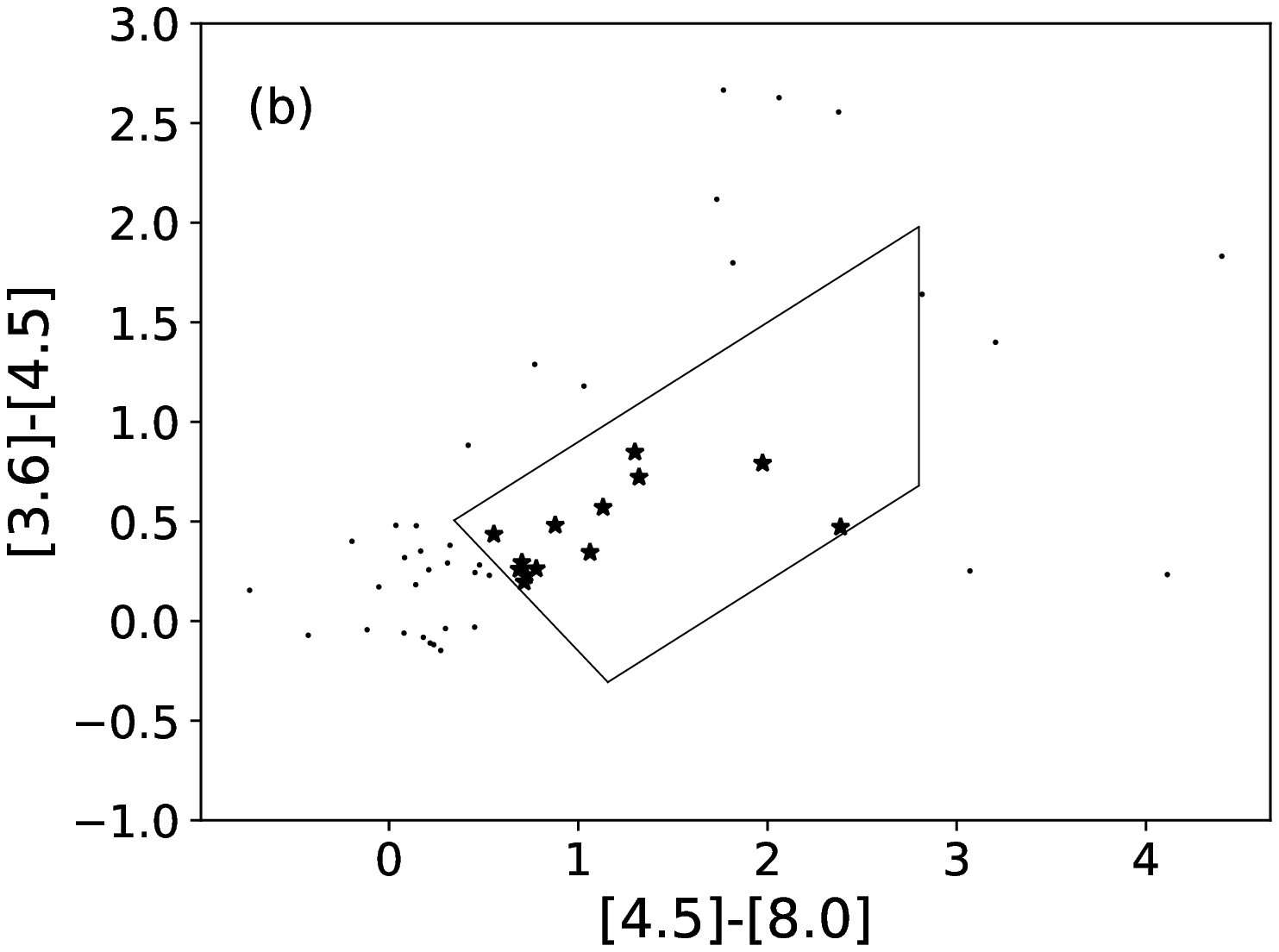}
                \includegraphics[scale=0.4]{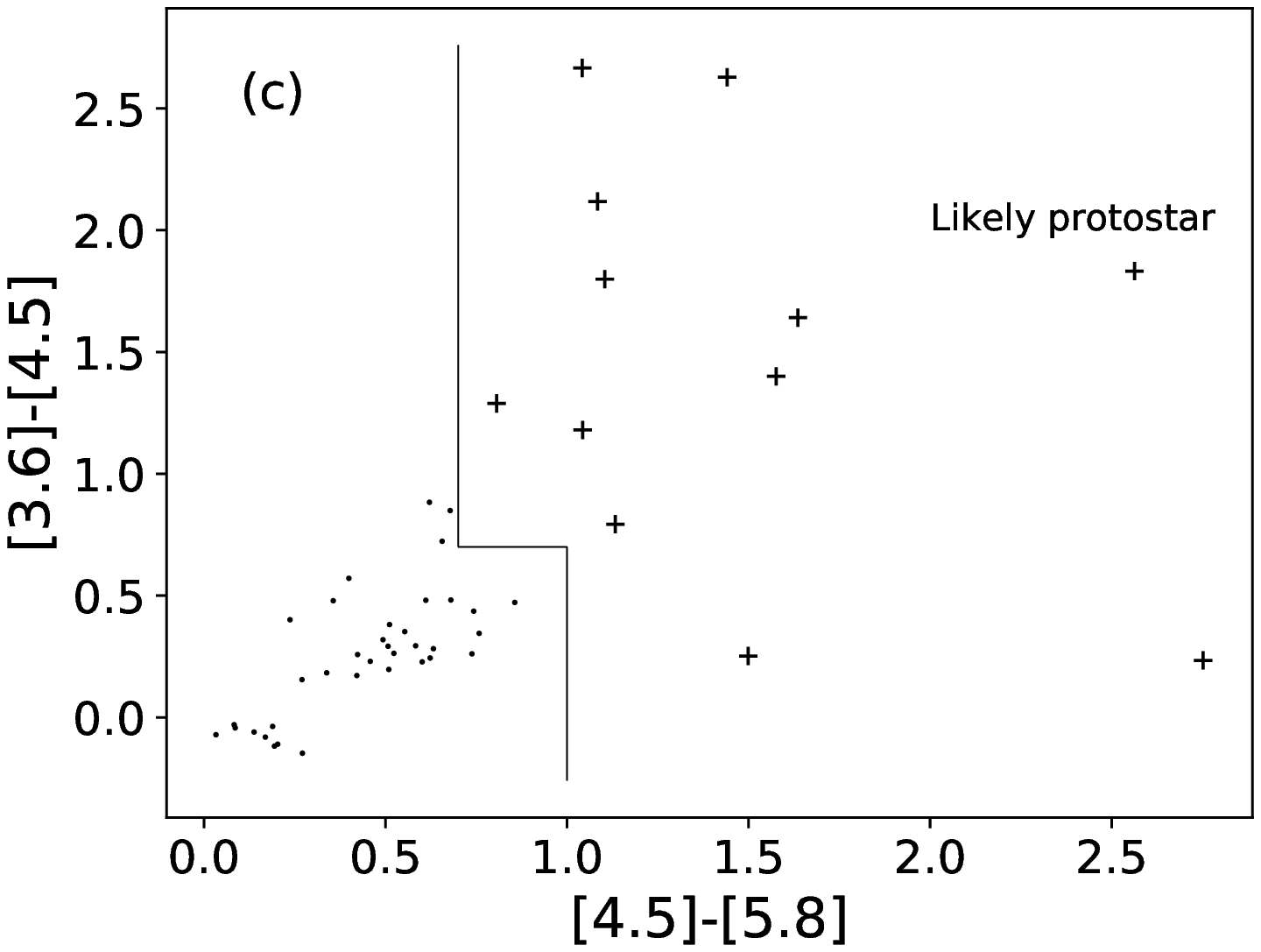}
                \includegraphics[scale=0.4]{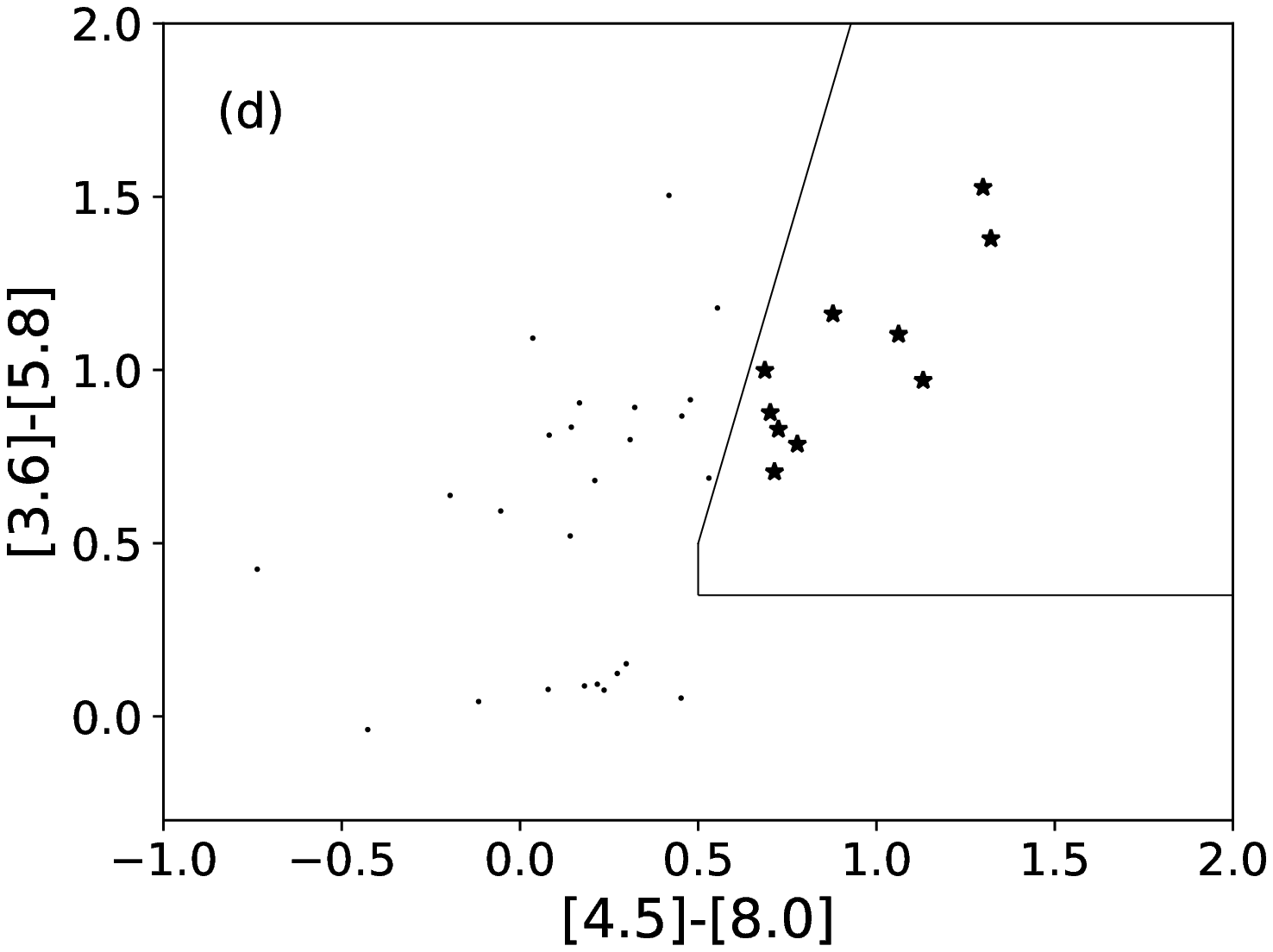}
                \includegraphics[scale=0.4]{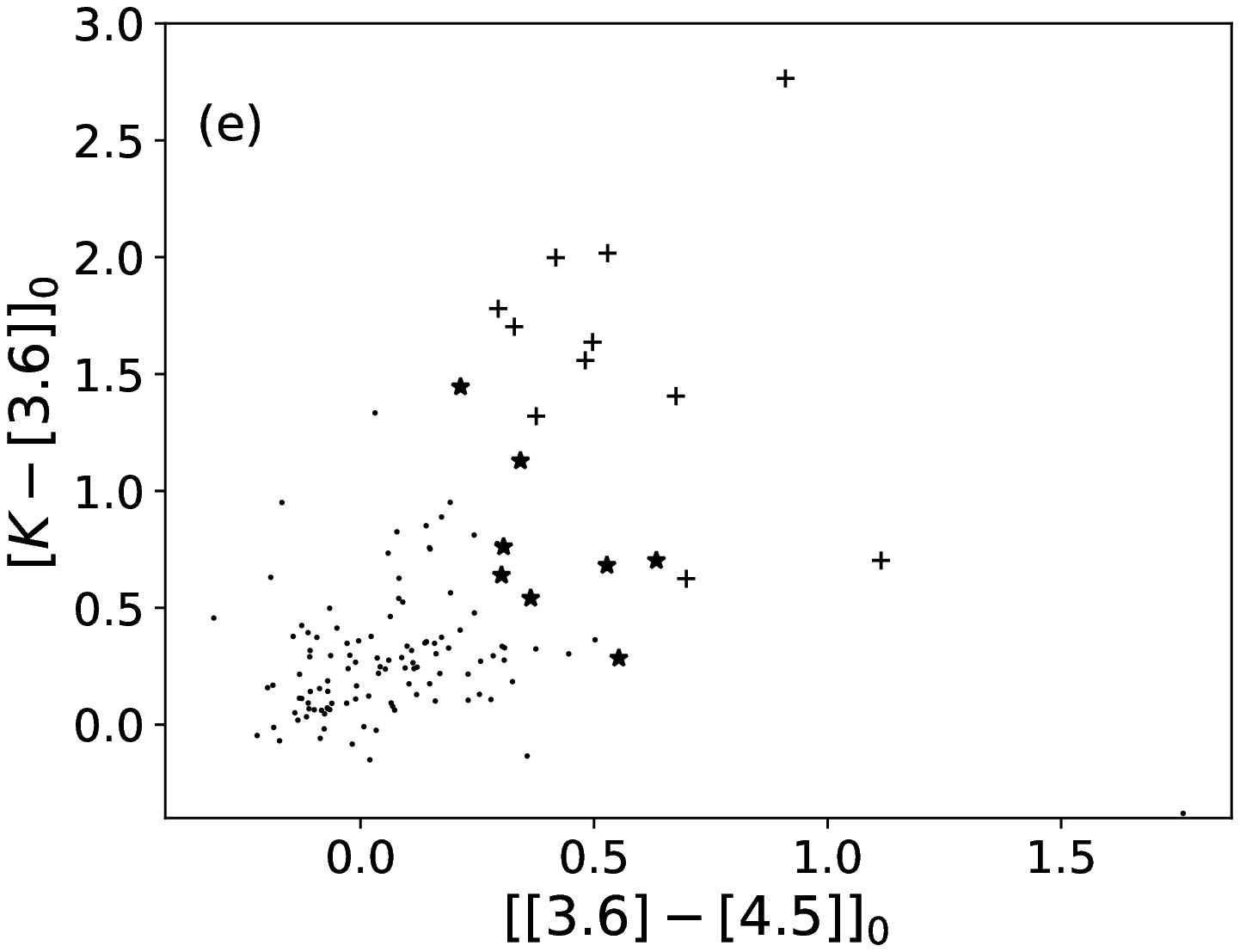}
                \includegraphics[scale=0.4]{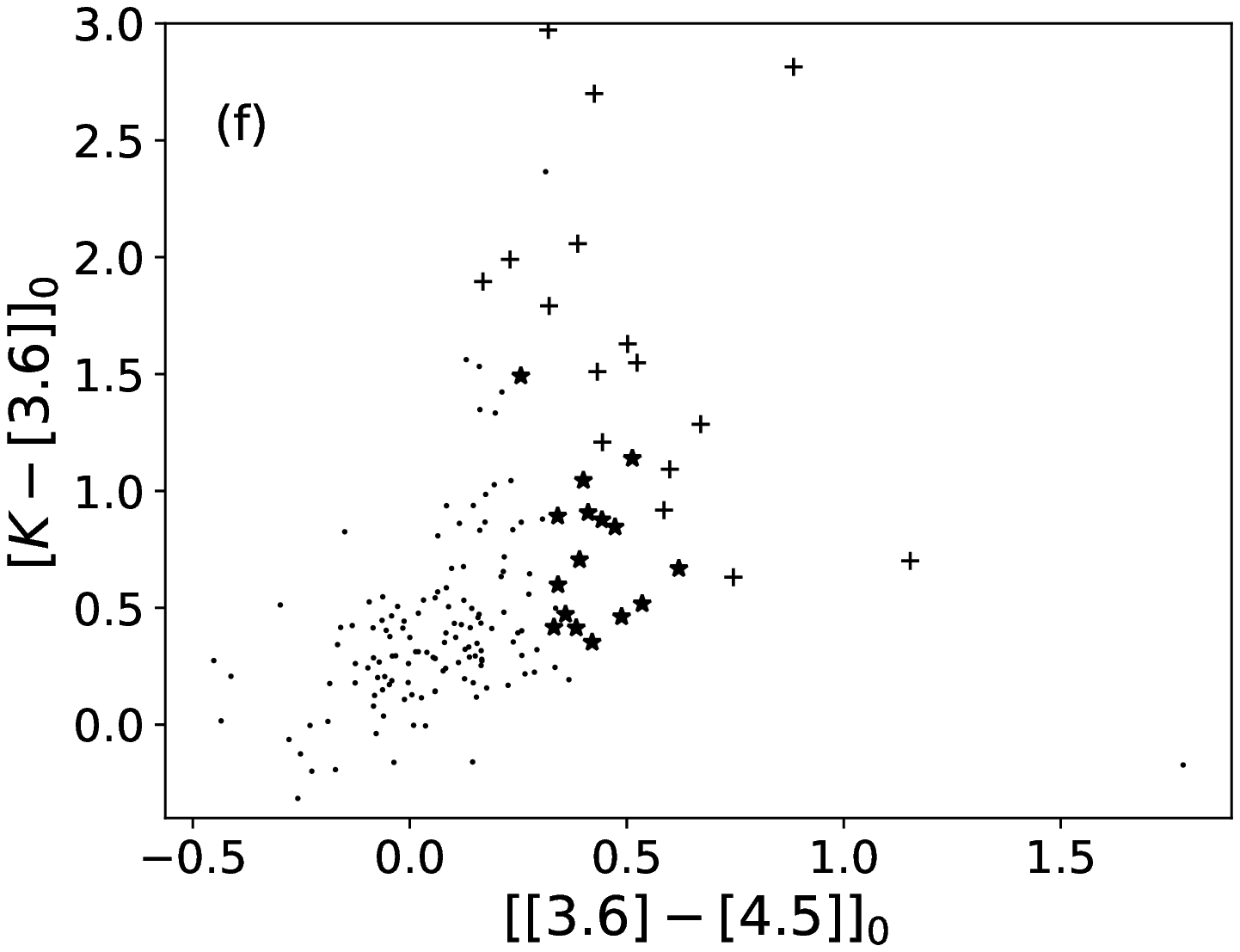}
                \caption{Identification of YSOs using color-color criteria of (a) \cite{allen:82}, (b)~\citet{simon:81}, (c) \citet{gutermuth:80} (Class I protostars), (d) \citet{gutermuth:80} (Class II YSOs), (e) \citet{gutermuth:80} with NIR data from 2MASS, and (f) \citet{gutermuth:80} with NIR data from UKIDSS. The cross, filled star, and filled triangles represent Class I, Class II, and Class I/II YSOs, respectively.}\label{fig:yso}
        \end{figure*}

        The YSO population in the W33 Main region has been identified using the mid- and near-infrared colors. We first searched for point sources in the GLIMPSE, 2MASS, and UKIDSS (Galactic Plane Survey) catalogs within a 5$\arcmin$ region centered at ($\alpha$, $\delta$) = (18$^h$14$^m$14$\fs$6, $-$17$\degr$55$\arcmin$47$\farcs$5). We then constructed color-color diagrams (Fig.~\ref{fig:yso}) and used the criteria mentioned in \citet{allen:82}, \citet{simon:81}, and \citet{gutermuth:80} to identify and classify YSOs. 
        
        We used the criteria of \citet{allen:82} in the $[3.6]-[4.5]$ versus $[5.8]-[8.0]$ color-color diagram (Fig.~\ref{fig:yso}(a)) to identify ten Class I and three Class I/II YSOs in the region of interest. \citet{simon:81} outlined criteria to identify YSOs based on the $[3.6]-[4.5]$ versus $[4.5]-[8.0]$ color-color diagram. This was used to identify 13 YSOs in the region (Fig.~\ref{fig:yso}(b)). \citet{gutermuth:80} used criteria on the $[3.6]-[4.5]$ and $[4.5]-[5.8]$ colors to identify Class I YSOs, and additional criteria based on $[3.6]-[4.5]$ and $[5.8]-[8.0]$ colors to detect Class II YSOs. Using these techniques, we identified 12 Class I and 11 Class II YSOs in the region (Fig.~\ref{fig:yso}(c) and (d)). Using all four IRAC bands, we identified a total of 13 Class I, eight Class II, and four Class I/II YSOs.
        
        The criteria above require photometry in all four IRAC bands for YSO identification. However, the GLIMPSE image shows the presence of significant extended emission at 8~$\mu$m in the vicinity of the radio continuum emission due to which photometry at 8~$\mu$m is often missing for sources that are detected in other IRAC bands. Hence, we used near-infrared photometry in $J$, $H$, and $K$ bands from 2MASS and UKIDSS surveys and determined the intrinsic $[K-[3.6]]$ and $[3.6]-[4.5]$ colors following the method given in \citet{gutermuth:80}. While the 2MASS counterparts to the GLIMPSE point sources are provided in the GLIMPSE catalog, the UKIDSS counterparts were identified using a matching radius of 1.2$''$. The UKIDSS colors and magnitudes were then converted to the 2MASS photometric system using the transformations of \citet{2001AJ....121.2851C}, after which the intrinsic colors were calculated. Using 2MASS and UKIDSS photometry bands, we identified 11 Class I and eight Class II YSOs, and 21 Class I and 16 Class II YSOs, respectively. The number of YSOs detected using the UKIDSS band is larger than the 2MASS band due to the higher sensitivity of the UKIDSS survey. We detect a total of 25 Class I and 23 Class II sources in the region using this method after cross-matching (Figure~\ref{fig:yso}(e) and (f)).

        \begin{figure}[h!]
                \centering
                \includegraphics[scale=0.5]{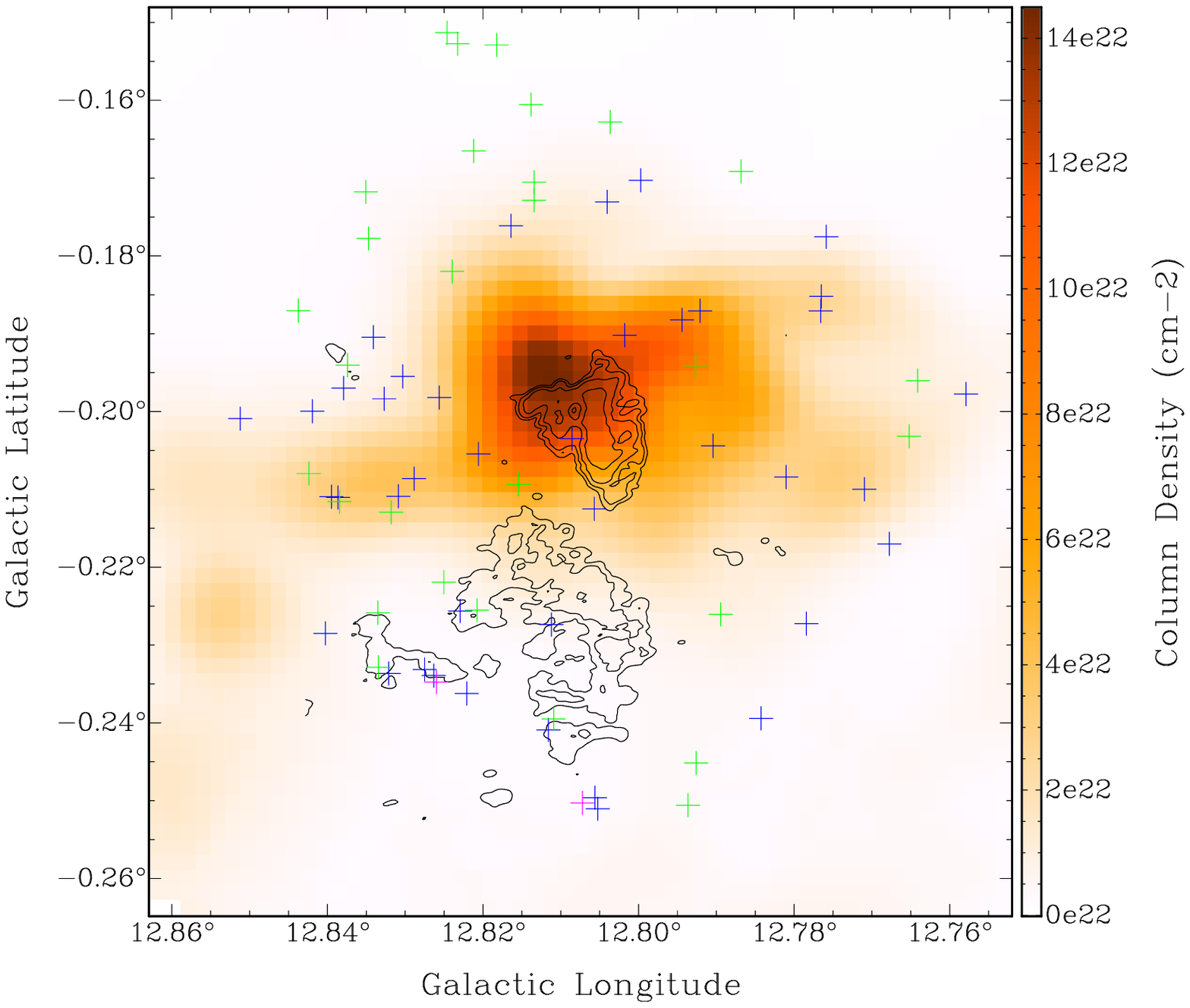}
                \caption{Locations of Class I (blue cross), Class II (red cross) and Class I/II (magenta cross) YSOs overlaid on the hydrogen column density map with the 1400~MHz black contours. The contour levels are as described in Figure \ref{fig:cont_map} (left panel).}\label{fig:yso_location}
        \end{figure}

        \begin{figure}[h!]
                \centering
                \includegraphics[scale=0.6]{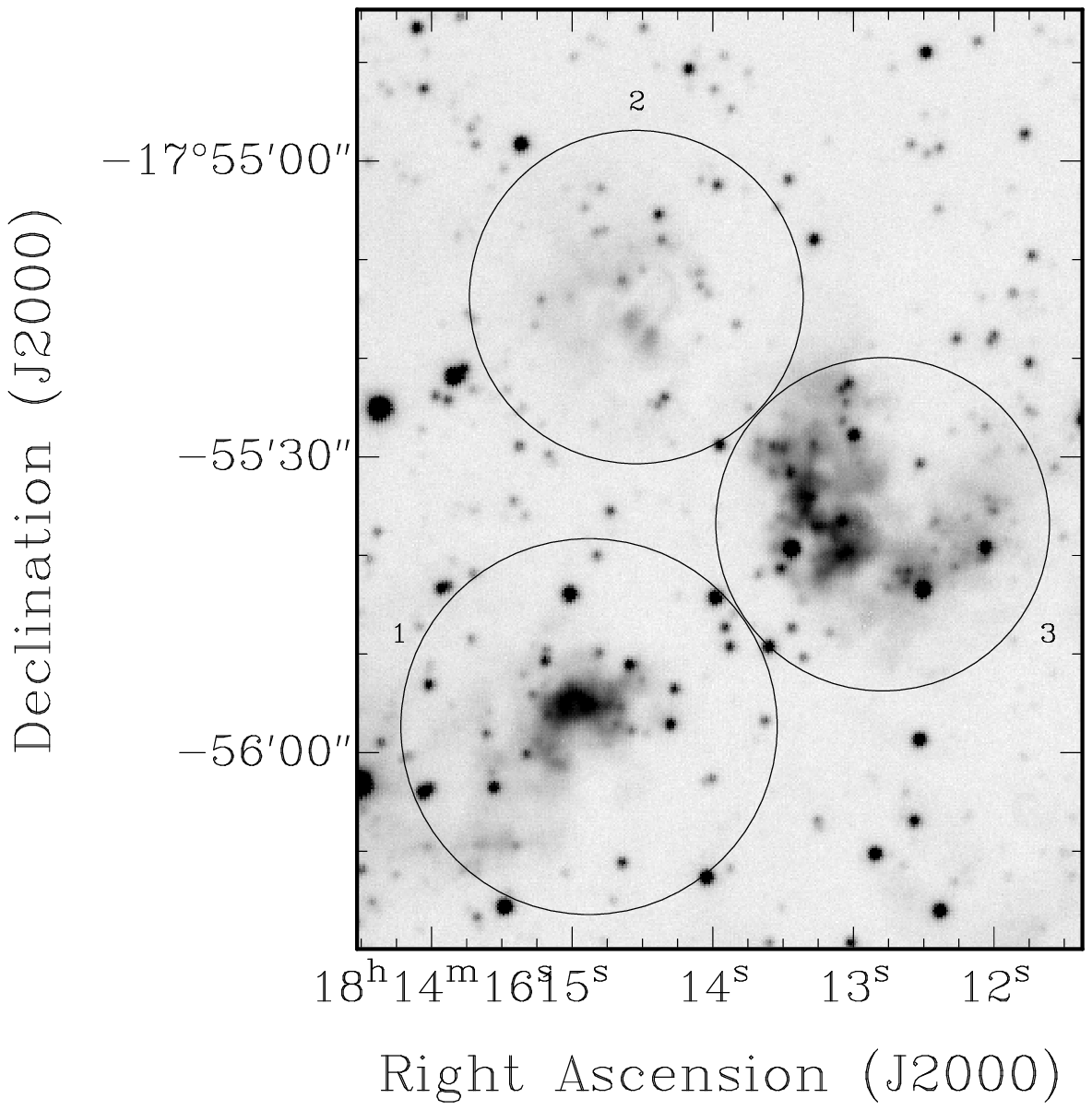}
                \caption{Gray-scale UKIDSS K-band image of W33 Main. The black circles represent the HII region residing inside W33 Main.}\label{fig:k_band}
        \end{figure}
        
        Combining the various detection techniques, we have identified a total of 40 Class I, 29 Class II, and two Class I/II sources in the region. Fig.~\ref{fig:yso_location} shows the distribution of identified YSOs on the map of the column density of molecular hydrogen. It is found that there is little correlation between the distribution of molecular gas and the distribution of YSOs, except for a small group of Class I YSOs to the west of W33 Main. This is mostly due to the significant extended emission associated with the HII region, which caused only very few point sources to be detected in the molecular clump. Moreover, the column density distribution of molecular hydrogen suggests that the extinction in the K band in the central region of W33 Main is as high as 7.5 magnitudes, which is equivalent to visual extinction of A$_V$ $\approx$ 67. The high extinction even in near- and mid-infrared wavelengths along with the presence of diffuse emission (Figure~\ref{fig:k_band}) allow for a significant undetected YSO population in W33 Main. The YSO population is seen to correlate with the large-scale cloud structure well in the W33 region \citep{2020MNRAS.496.1278D, 2021A&A...646A.137L}.
        
        \subsection{Nature of G12.81$-$0.22}\label{sec:nature_of_g12}
        The spectral index of radio emission in G12.81$-$0.22 shows that the emission is optically thin thermal emission. This together with the arc-like morphology of the source suggests that it traces part of an older HII region. The size and emission measure of G12.81$-$0.22 (Table~\ref{tab:physicalpara}) also show it to be consistent with a compact HII region. The radial velocity gradient across the arc suggests that the HII region is expanding with a velocity $\sim 6$~km~s$^{-1}$. The Str\"omgren radius ($R_{\text{st}}$) of HII region is given by the following \citep{Tielens}:
        \begin{equation}
        R_{\text{st}} \simeq 1.2\left(\frac{10^3 \text{cm}^{-3}}{n_{\text{H}_2}}\right)^{2/3}\left(\frac{N_{\text{Ly}}}{5 \times 10^{49}}\right)^{1/3} \text{pc}
        ,\end{equation}
        where $N_{\text{Ly}}$ is the Lyman continuum photon rate and $n_{\text{H}_2}$ is the mean hydrogen density. Assuming that G12.81$-$0.22 had an initial hydrogen density that is characteristic of what is observed toward W33 Main at the present time, the Str\"omgren radius (R$_{st}$) of the region is 0.03~pc, which is much smaller than the current size of the HII region ($\sim 0.4$~pc). This, together with the low continuum optical depth even at 610 MHz, shows that G12.81-0.22 is a much more evolved HII region and it is expanding on account of the pressure difference between the HII region and the ambient medium. Furthermore, this expansion is likely to be slowing down as mentioned in \S\ref{sec:gasdynamics}. Assuming expansion into a homogeneous molecular cloud, the dynamical age of the HII region was estimated using following equation~\citep{wood:7}:
        \begin{equation}
        t_{\text{dyn}} = \frac{4}{7}\frac{R_{\text{st}}}{c_s}\left[\left(\frac{R_{f}}{R_{\text{st}}}\right)^{7/4}-1\right]
        ,\end{equation}
        where $R_f$ is the final radius of the HII region and $c_s$ is the isothermal sound speed. Taking the final radius as 0.4~pc and adopting a sound speed of 0.5~km~s$^{-1}$ (corresponding to a mean cold dust temperature of 35~K), we estimate the dynamical age of G12.81$-$0.22 to be 3.1~Myr. This is consistent with the stellar age $\sim 2-4$~Myr of the massive stars in the W33 massive star-forming region \citep{messineo}.
        
        \section{Summary}
        
        We have carried out a multiwavelength study of the W33 Main region. High-resolution observations at 1400 and 610~MHz, using the GMRT and 4$-$8~GHz data from GLOSTAR, reveal the presence of two radio continuum sources in the region. The HII region associated with W33 Main is found to be moderately optically thick at 4$-$8~GHz and optically thick at 1.4~GHz, while the diffuse continuum source G12.81$-$0.22 is optically thin at all observed frequencies. Radio recombination lines are detected toward both continuum sources at a good S/N in GLOSTAR, while the H167$\alpha$ line was marginally detected by GMRT. The GLOSTAR recombination line data show the presence of a velocity gradient across W33 Main. Our work shows that the HII regions toward W33 Main are expanding and interacting with the surrounding material. The peak dust temperature in the region was found to be 45~K which is consistent with W33 Main being a relatively evolved massive star-forming region. While a total of 71 YSOs were detected in the region, the presence of extended emission and high extinction at K band allows for the presence of a significant population of undetected YSOs that would be directly associated with the W33 Main molecular cloud. The physical properties of W33 Main reflect the evolved stage of the region and the properties of the diffuse continuum source, G12.81$-$0.22, the dynamics of ionized gas, and its dynamical age are consistent with the stellar age of the W33 complex.
        
        \begin{acknowledgements} 
                We thank the staff of the GMRT that made these observations possible. GMRT is run by the National Center for Radio Astrophysics (NCRA) of the Tata Institute of Fundamental Research. DVL acknowledges the support of the Department of Atomic Energy, Government of India, under project no. 12-R\&D-TFR-5.02-0700.
        \end{acknowledgements}

        \bibliographystyle{aa} % style aa.bst
        \bibliography{mybib.bib} % your references Yourfile.bib
        
        \begin{appendix}
                \section{Derivation of electron temperature}\label{sec:dev_te}
                The detection of the recombination line allowed us to determine the electron temperature of the region using the line-to-continuum ratio using the following equations:
                \begin{equation}
                \label{eq:tau_c}
                \tau_C=-\ln\left[1-\frac{c^2I_C}{2kT_e\nu^2}\right]
                \end{equation}
                \begin{equation}
                \label{eq:tau_l}
                \tau_L=-\ln\left[1-\frac{I_L}{I_C}\exp\left(\tau_C\right)\left(1-\exp\left(-\tau_C\right)\right)\right]
                ,\end{equation}
                where $I_C$ and $I_L$ refer to the specific intensity of the continuum and line, respectively, and $\tau_C$ and $\tau_L$ are the continuum and line optical depth, respectively. The optical depth $\tau_L$ corresponds to the center frequency of the line in terms of the emission measure given by the following~\citep{2009tra..book.....W}:
                \begin{equation}
                \label{eq:tau_l_n}
                \tau_L \approx 1.92\times 10^3 \left( \frac{T_e}{\text{K}}\right)^{-5/2}\left(\frac{\triangle\nu}{\text{kHz}} \right)^{-1}\left(\frac{EM}{\text{pc cm}^{-6}} \right)  
                ,\end{equation}
                where $\Delta \nu$ is the line width. Under the Altenhoff approximation \citep{altenhoff1960veroff}, the continuum optical depth is given by
                \begin{equation}
                \label{eq:tau_c_n}
                \tau_C \approx 8.24 \times 10^{-2} \left( \frac{T_e}{\text{K}}\right)^{-1.35}\left(\frac{\nu}{\text{GHz}} \right)^{-2.1}\left(\frac{EM}{\text{pc cm}^{-6}} \right)
                .\end{equation}
                When the radio continuum has a moderate optical depth, the electron temperature (T$_e$) can thus be obtained from the ratio of line-to-continuum optical depth:   
                \begin{equation}
                \label{eqn:final}
                T_e = \left[2.33 \times 10^4 \left(\frac{\nu}{\text{GHz}} \right)^{2.1} \left(\frac{\text{kHz}}{\triangle \nu} \right) \frac{\tau_C}{\tau_L}  \right]^{1/1.15} 
                .\end{equation}
                
                When the radio continuum is optically thin, one can approximate $\tau_C/\tau_L$ to be $I_C/I_L$ in eq.~(\ref{eqn:final}) and it can estimate the electron temperature using the following equation (see, eg.,~\cite{2019ApJ...887..114W}):
                \begin{equation}
                \label{eq:op_thin_intial}
                \left(\frac{T_e}{\text{K}}\right)=\left\{3.661\times 10^4 \frac{\Delta n}{n}f_{nm}\left(\frac{\nu}{\text{GHz}}\right)^{1.1}\left(\frac{I_C}{I_L}\right)\left(\frac{\Delta v}{\text{km s}^{-1}}\right)^{-1}\left[1+\frac{n(^4\text{He}^{+})}{n(\text{H}^{+})}\right]^{-1}\right\}^{0.87}
                ,\end{equation}
                
                where the expression $f_{nm}$ is the absorption oscillator strength and its approximation is given by~\citep{ 1968Natur.218..756M}
                \begin{equation}
                \label{eq:oscillator}
                f_mn = nM_{\Delta n}\left(1+1.5\frac{\Delta n}{n} \right) 
                ;\end{equation}          
                
                the expression ($\Delta n/nf_{nm}$) = 0.19345 for $\Delta n = 1$ and $n$=107. After substitution, the final expression for the electron temperature is given by                
                \begin{equation}
                \label{eq:op_thin}
                \left(\frac{T_e}{\text{K}}\right)=\left\{7082.2\left(\frac{\nu}{\text{GHz}}\right)^{1.1}\left(\frac{I_C}{I_L}\right)\left(\frac{\Delta v}{\text{km s}^{-1}}\right)^{-1}\left[1+\frac{n(^4\text{He}^{+})}{n(\text{H}^{+})}\right]^{-1}\right\}^{0.87}
                ,\end{equation}
                where $\Delta v$ is the line width and $n(^4\text{He}^{+})/n(\text{H}^{+})$ is the ionic abundance ratio, which is considered to be a constant value of 0.07 $\pm$ 0.02~\citep{quireza:62}.        
                
                When the optical depth was moderate, an iterative procedure was followed, wherein an initial guess for $T_e$ was provided to obtain the values of $\tau_C$ and $\tau_L$ from eqs. (\ref{eq:tau_c}) and (\ref{eq:tau_l}) after which a revised estimate of $T_e$ was obtained from eq.~(\ref{eqn:final}). The process was repeated until $T_e$ converged.%\ \LEt{ Single-sentence paragraphs are not allowed, thus the proposed edit.}The process was repeated until $T_e$ converged.
                
                \section{Physical parameters of HII regions}\label{sec:phy_pro}
                The electron density and Lyman continuum flux rate at 5.8~GHz can be estimated using the equation given in \citet{2016A&A...588A.143S}:
                \begin{equation}
                \label{eq:n_e}
                \left(\frac{n_{e}}{\text{cm}^{-3}}\right) = 2.576\times 10^6\left(\frac{F_\nu}{\text{Jy}}\right)^{0.5}\left(\frac{T_e}{\text{K}}\right)^{0.175}\left(\frac{\nu}{\text{GHz}}\right)\times \left(\frac{\theta_{source}}{\text{arcsec}}\right)^{-1.5}\left(\frac{D}{\text{pc}}\right)^{-0.5}
                \end{equation}
                
                \begin{equation}
                \label{eq:n_ly}
                \left(\frac{N_{Ly}}{\text{s}^{-1}}\right) = 4.771\times 10^{42}\left(\frac{F_\nu}{\text{Jy}}\right)\left(\frac{T_e}{\text{K}}\right)^{-0.45}\left(\frac{\nu}{\text{GHz}}\right)^{0.1}\left(\frac{D}{\text{pc}}\right)^{2}
                ,\end{equation}
                where $\theta_{source}$ is the source size taken from the catalog of~\citet{Anderson_2014}, F$_\nu$ is the flux density at frequency $\nu$, and D is the distance to the source.\\
                Furthermore, the corresponding mass of ionized gas was estimated using the following equation~\citep{Tielens}:
                \begin{equation}
                \label{eq:mass}
                M_\sun = 80 \times \left(\frac{10^3\text{cm}^{-3}}{n_e}\right)\left(\frac{N_{Ly}}{5\times 10^{49}\text{s}^{-1}}\right)
                .\end{equation}

        \end{appendix}
        
\end{document}